\pdfoutput=1
\documentclass[printer]{aa}

\usepackage{txfonts}
\usepackage{graphicx}
\usepackage[tight,scriptsize]{subfigure}
\usepackage{epstopdf}
\usepackage{float}

\usepackage[process= auto, crop=pdfcrop]{pstool}
\usepackage{import}

\usepackage{natbib}

\newcommand{\Mm}{\,\mathrm{Mm}}
\newcommand{\minutes}{\,\mathrm{min}}
\newcommand{\hr}{\,\mathrm{hr}}
\newcommand{\simm}{\mathtt{\sim}}
\newcommand{\arcs}{\,\mathrm{arcsec}}

\newcommand{\px}{\,\mathrm{px}}
\newcommand{\s}{\,\mathrm{s}}
\newcommand{\G}{\,\mathrm{G}}
\newcommand{\DNS}{\,\mathrm{DN\,s^{-1}}}
\newcommand{\Mx}{\,\mathrm{Mx}}

\newcommand{\Tdh}{T_{dh}}
\newcommand{\Tdz}{T_{dz}}

\newcommand{\Fig}[1]{Fig.\,\ref{#1}}
\newcommand{\Figs}[1]{Figs.\,\ref{#1}}
\newcommand{\Eq}[1]{Eq.\,\ref{#1}}

\newcommand{\Section}[1]{Sect.\,\ref{#1}}
\newcommand{\Sections}[1]{Sects.\,\ref{#1}}

\begin{document}
\title{Magnetic balltracking: Tracking the photospheric magnetic flux}

\author{R. Attie  
\and D. E. Innes}
\institute{Max-Planck-Institut f\"{u}r Sonnensystemforschung, 37077 G\"ottingen, Germany}

\abstract
{One aspect of understanding the dynamics of the quiet Sun is to quantify the evolution of the flux within small-scale magnetic features. These features are routinely observed in the quiet photosphere and were given various names, such as pores, knots, magnetic patches.}
{This work presents a new algorithm for tracking the evolution of the broad variety of small-scale magnetic features in the photosphere, with a precision equal to the instrumental resolution.}
{We have developed a new technique to track the evolution of the individual magnetic features from magnetograms, called "magnetic balltracking". It quantifies the flux of the tracked features, and it can track the footpoints of magnetic field lines inferred from magnetic field extrapolation. The algorithm can detect and quantify flux emergence, as well as flux cancellation.}
{The capabilities of magnetic balltracking are demonstrated with the detection and the tracking of two cases of magnetic flux emergence that lead to the brightening of X-ray loops. The maximum emerged flux ranges from $10^{18} \Mx$ to $10^{19} \Mx$ (unsigned flux) when the X-ray loops are observed.}
{}


\maketitle

\section{Introduction}

Describing the evolution of the magnetic flux on the surface of the Sun is a key component for understanding the complex couplings involved in energetic events that release a considerable amount of energy into the solar atmosphere and beyond. For the quiet photosphere, tracking algorithms already exist that describe the motion and some physical properties of the magnetic features. Each of them have different limitations. Because of that, the most complete description of the photospheric magnetic flux cannot be achieved by using only one of these algorithms. Instead, combining different algorithms with different limitations, which are therefore complementary, gives a more complete description of the behaviour of the photospheric magnetic features. 

Among the many algorithms that attempt to analyse the magnetic field,  a consistent set of four algorithms have been developed and compared with each other in \citet{DeForest2007}, where they are referred to as CURV \citep{Hagenaar1999}, MCAT \citep{Parnell2002}, YAFTA \citep{Welsch03,DeForest2007,Stangalini2014}, and SWAMIS \citep{DeForest2007,Lamb2008}. 
Other algorithms designed to track magnetic features exist, such as those described in \citet{Demoulin2003} and \citet{Welsch2004}, and they focus on deriving averaged velocity flows in magnetised regions. 

The present paper describes a new algorithm using a completely different paradigm than the algorithms mentioned above. It is also designed to track and label small magnetic features in the quiet photosphere, as small as can be seen, and it measures the magnetic flux they carry. This algorithm does not pretend to replace these well-tested algorithms, but it aims at providing another alternative that simply broadens the spectrum of science applications that can benefit from feature tracking techniques in magnetograms, more specifically at the smallest scales (a few $\Mm$ or less).

This technique uses the same paradigm as the one used in a known algorithm called balltracking \citep{Potts04,Attie09}, which is designed to track granules in continuum images. In this technique, numerical spheres or "balls" are released onto the image. Their position is known at any given time, and they settle in the local intensity minima that move with the granules so we can track the group motions of the latter. However, because of the very different nature of the data provided by magnetograms, we needed to write a variant of this algorithm, which we refer to as  "magnetic balltracking". 

After a brief summary of the balltracking paradigm in \Section{sec:balltracking}, the magnetic balltracking is explained in four main phases in \Section{sec:phases}. In \Sections{sec:rescaling} and \ref{Radius} we discuss a few technical aspects of the algorithm. Its practical application is detailed in \Section{sec:regiongrow} and applied in \Section{sec:flux_emergence} to a case study of flux emergence  associated with the brightening of X-ray loops.

\section{The balltracking paradigm \label{sec:balltracking}}

\begin{figure*}
	\centering
	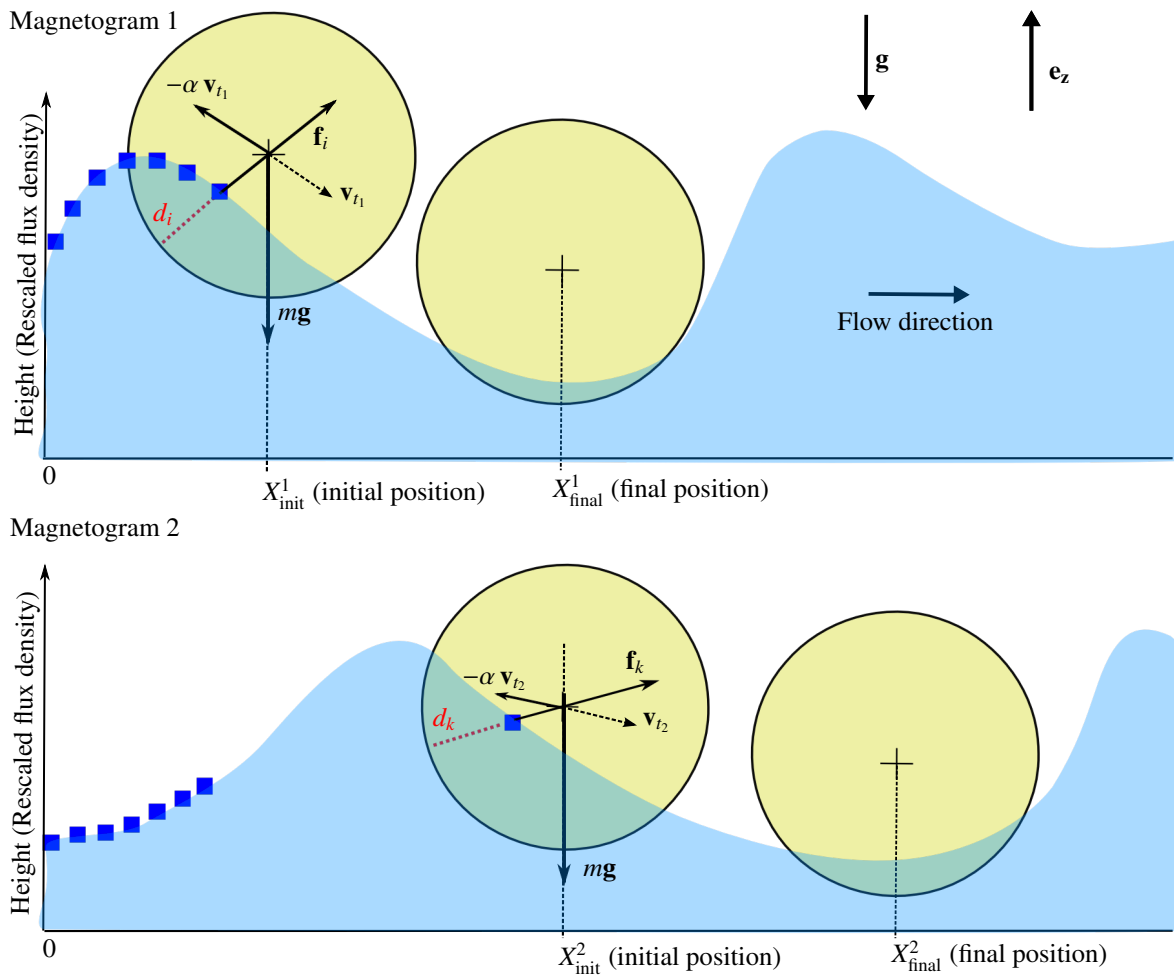 
	\caption{General concept of the balltracking algorithm for one ball. The blue squares represent the pixels drawn at a height equal to the rescaled value of the unsigned flux density so that maxima in flux density appear as local minima.  Only a few pixels are represented.}
	\label{balltrack_mag}
\end{figure*}

The full derivation of the equations of motion that animate the balls within balltracking can be found in \citet{Potts04}. Here we schematise the balltracking paradigm in the context of tracking magnetic features. 
The general principle is sketched in Fig.\ref{balltrack_mag}, where the forces and the integration of the position of one ball is represented across two consecutive data surfaces at the times $t_1$ and $t_2$. These data surfaces are, in fact, pre-processed magnetograms (more on this in \Section{sec:mphase1}).

The three forces governing the motion of each ball can simply be expressed with the vectorial equation (vectors noted in bold letters):

\begin{equation}
m\mathbf{\dot{v}}=\sum\limits_{i}\mathbf{f}_i(d_i) - mg\mathbf{e_z} - \alpha \mathbf{v}
\label{ball_eq1}
\end{equation}
where $\mathbf{v}$ is the velocity of the ball and $\mathbf{\dot{v}}$ its time derivative. Here, $\sum\limits_{i}\mathbf{f}_i(d_i)$ is the sum over  all the elementary buoyancy forces exerted by the data point at the $i^{th}$ pixel at the penetration depth $d_i$. The user-defined gravity field $g$ sets the maximum acceleration that a ball can possibly reach. It is oriented downward, in a 3D cartesian frame of reference where $\mathbf{e_z}$ is a unit vector pointing upward and $\alpha$ is a damping coefficient that controls the stability of the tracking. For simplicity, the mass $m$ of the balls can be set to unity. In \Fig{balltrack_mag}, the ball is moving with a magnetic feature, where the intensity of the magnetic flux density is reversed and rescaled so it can be seen as a geometric height. In such a picture, the local maximum of unsigned flux density is, in fact, a local minimum. While floating on the data surface represented by the blue waves, the ball settles in the local minimum, while the latter moves with the group motion.
In this framework, we are only interested in using the final position of the balls at the end of the integration of \Eq{ball_eq1} in each magnetogram.

\section{The magnetic balltracking algorithm \label{sec:phases}}

\subsection{Phase 1: preprocessing of the magnetograms \label{sec:mphase1}}
\begin{figure}
	\centering
	\subfigure{\includegraphics[width=1\columnwidth]{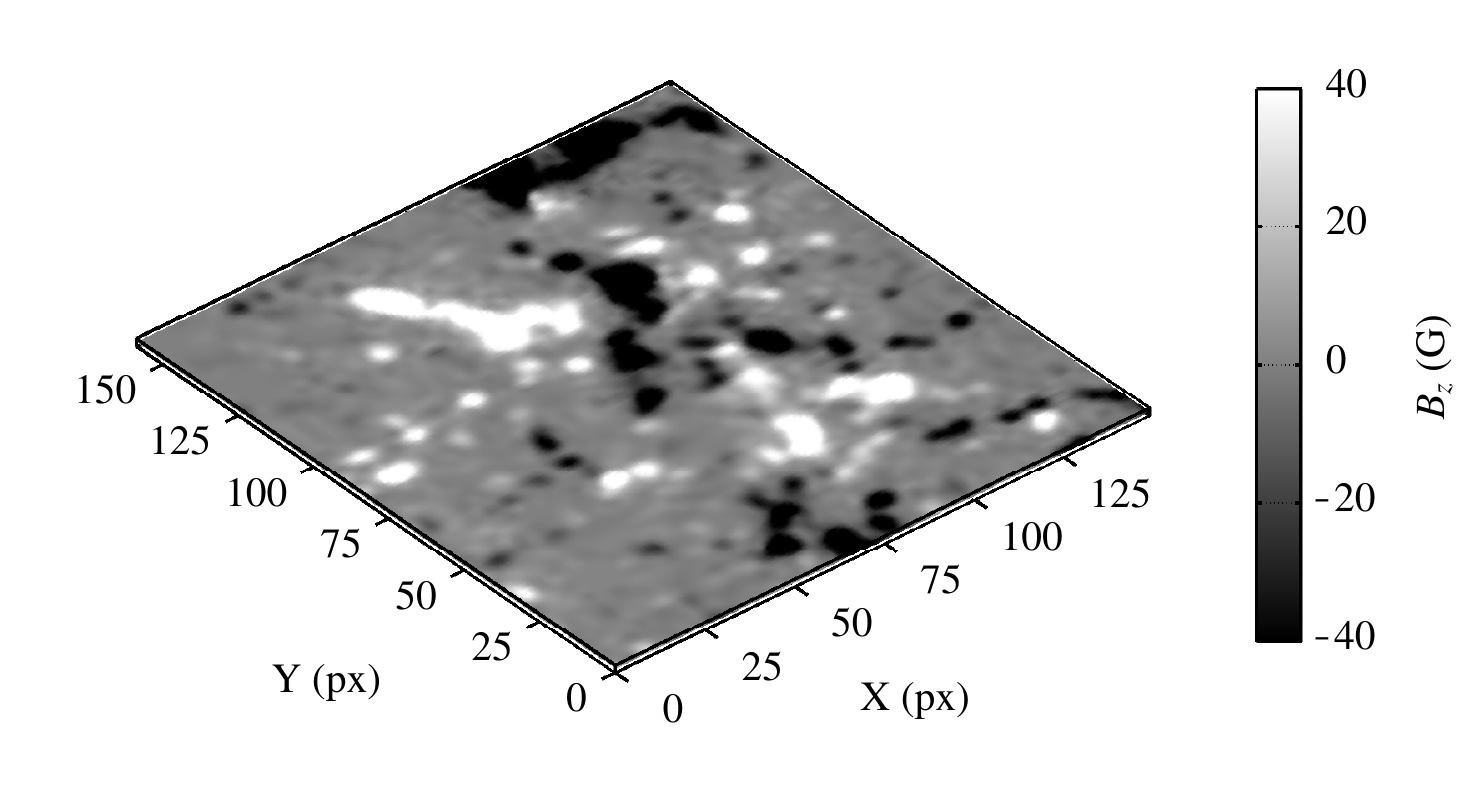}}\\
	\subfigure{\includegraphics[width=1\columnwidth]{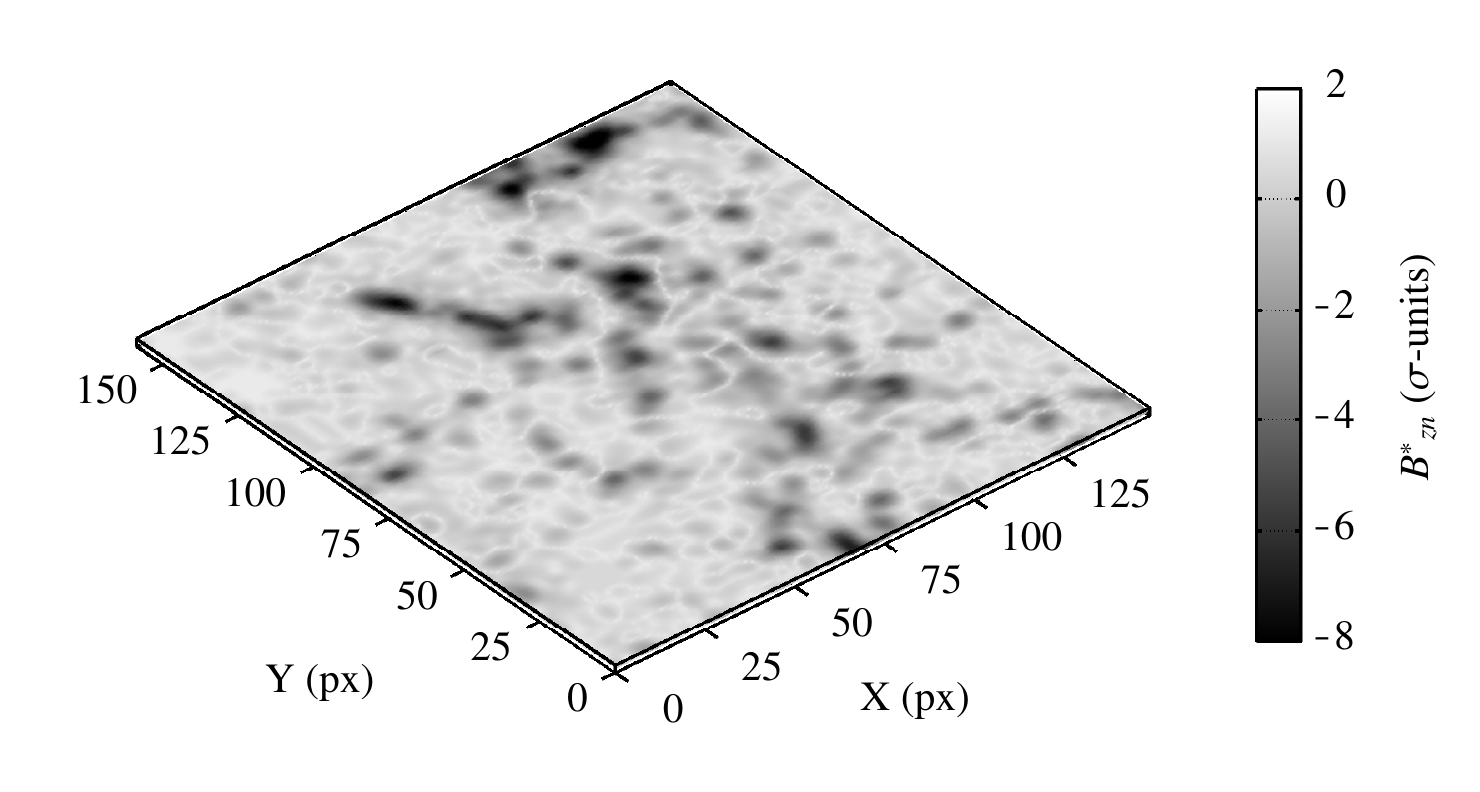}}\\
	\caption{Preprocessing of the magnetic balltracking. Top: Initial magnetogram $B_z$, calibrated into Gauss units, scaled between $-40\G$ and $+40\G$.
	Bottom: ${{B_{zn}}^{\star}}$ obtained after rescaling $B_z$ \label{mballtrack}} 
\end{figure}

\begin{figure}
	\centering
	\subfigure{\includegraphics[width=1\columnwidth]{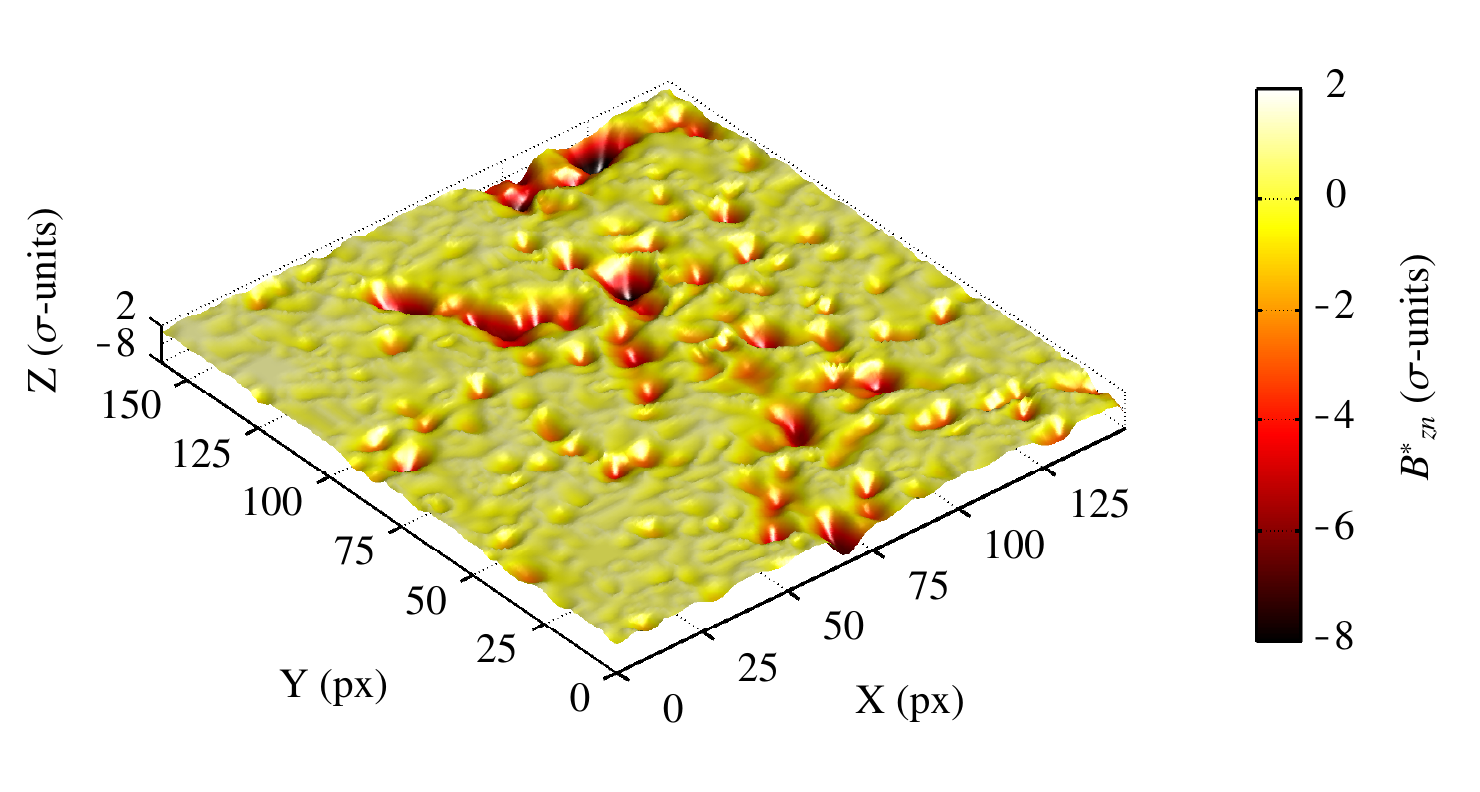}}\\
	\subfigure{\includegraphics[width=1\columnwidth]{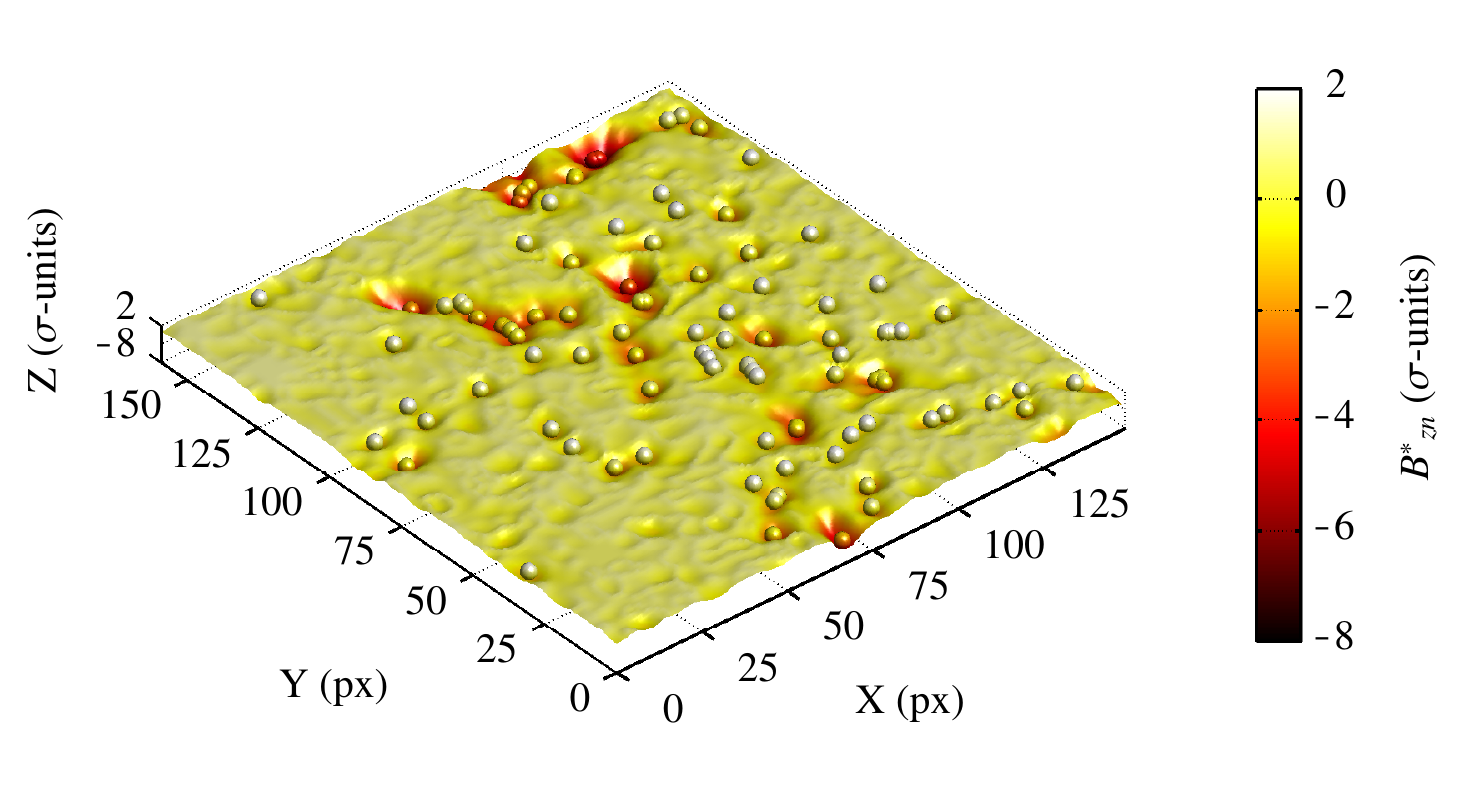}}\\	
	\caption{Top: ${{B_{zn}}^*}$ in 3D, using the same data as in the bottom of \Fig{mballtrack}, with the intensity used as a geometrical height. The colormap is scaled exactly as the intensity. Bottom: same as the top figure, with balls that have settled in the "magnetic holes" after a few integration steps. The position of these balls on the original 2D magnetograms are shown in \Fig{mballtrack_real_init}. \label{mballtrack3D}} 
\end{figure}

The first step of magnetic balltracking is to make the magnetic features "trackable" by the balls. The main change from the orignal balltracking is to account for the signed values of the magnetograms, and for the more contrasted maps, spanning typically over more than 2 orders of magnitude in the quiet Sun (from a few $\G$ to hundreds of $\G$). So we have to rescale the magnetograms in order to rescale the magnetic features vertically into "holes" of reasonable depth, from either positively or negatively signed magnetic flux density, and allow the balls to settle inside them. This is the purpose of the following preprocessing (see also \Figs{mballtrack} and \ref{mballtrack3D}).

To start with, let us consider the absolute value of the flux density in a single magnetogram as a scalar field $|B_z(x,y)|$, like the one in \Fig{mballtrack} (top). It is first reverse-scaled (non-linearly) into
\begin{equation*}
 B^*_z(x,y)=\mathrm{max}(\sqrt{|B_z(x,y)|})-\sqrt{|B_z(x,y)|}
\end{equation*}
Next, ${B^*_z}$ is offset by its mean value and normalised to its standard deviation $\sigma_{B^*_z}$ taken across the whole frame: 
\begin{equation*}
B^*_{zn}(x,y)=\frac{\displaystyle B^*_z(x,y)-<B^*_z(x,y)>}{\displaystyle \sigma_{B^*_z}} 
\end{equation*}
where $B^*_{zn}$ is displayed in \Fig{mballtrack} (bottom), with an intensity spanning over a few units, which is of the order of the horizontal size of the magnetic features that we want to track. 
This rescaled signal can be seen as a geometrical height. In a 3D plot (\Fig{mballtrack3D}), the magnetic features look like holes into which the balls can settle easily. This particular choice of rescaling (against for instance a linear rescaling) is further justified in \Section{sec:rescaling}.

\subsection{Phase 2: Initialisation  \label{sec:mphase2}}
\begin{figure}
	\centering
	\subfigure{\includegraphics[width=1\columnwidth]{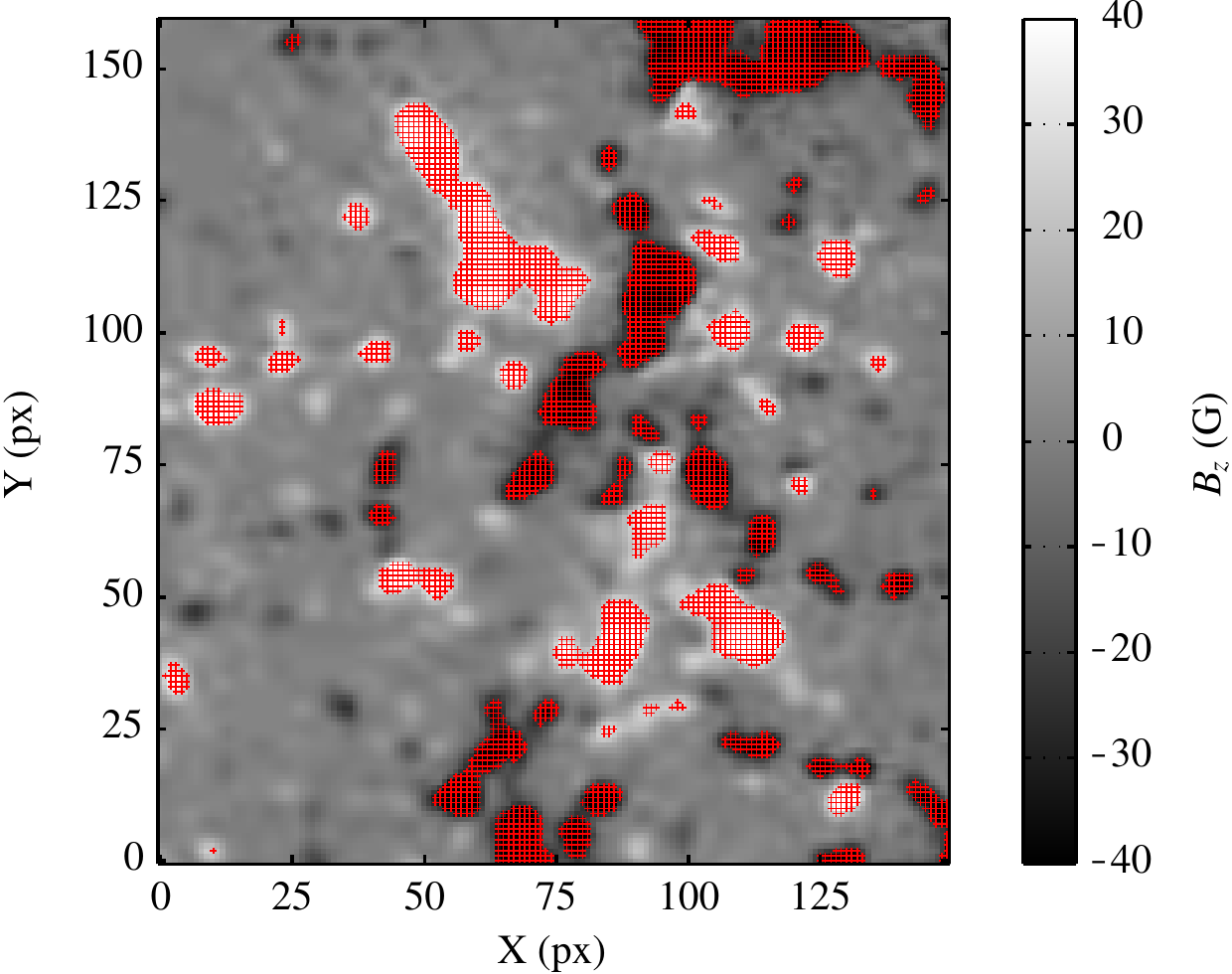}}\\
	\subfigure{\includegraphics[width=1\columnwidth]{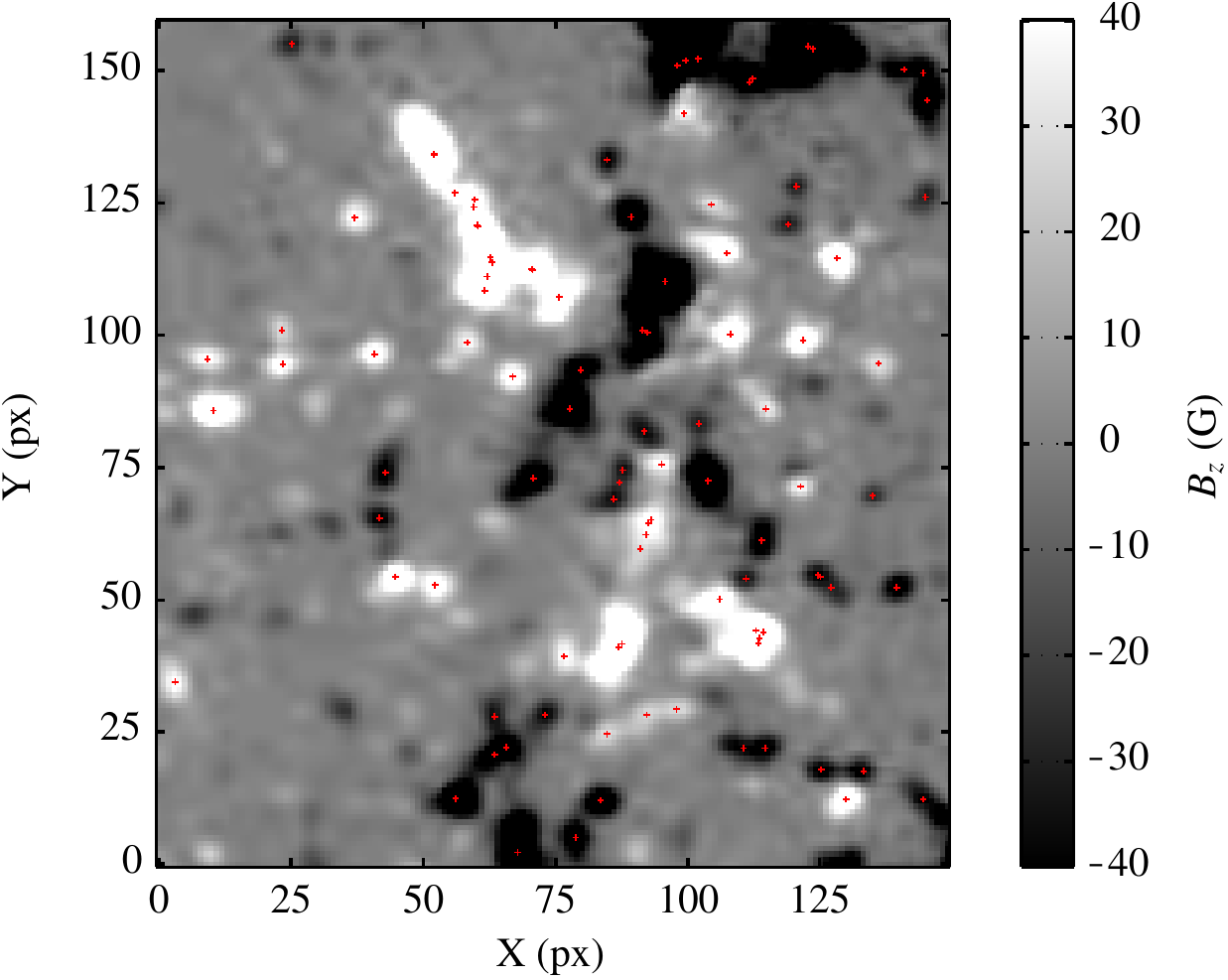}}
	\caption{Initialisation of the magnetic balltracking. Top: The values in the original magnetogram $B_z$ are used to dispatch the balls on pixels above $20\G$ (red crosses). Bottom: New positions of the balls, after integrating the equations of motion, on the same magnetogram.  \label{mballtrack_real_init}}
\end{figure}

Once the magnetograms are rescaled, the balls are initially positioned at the pixels whose absolute intensity in the original (not rescaled) magnetogram is above a given threshold. This initialisation is illustrated in \Fig{mballtrack_real_init} (top) where the balls' centres are plotted on the original magnetograms. 

With the magnetograms from the Narrowband Filter Imager of the Solar Optical Telescope (NFI/SOT) onboard Hinode \citep{Tsuneta08}, the threshold is usually set between $5\G$ and $20\G$. Regardless, these values should be chosen so one does not track random noise. In addition, this saves some computational time by reducing the number of balls that is much smaller than the total number of pixels. In the magnetic balltracking, we do not make assumptions on the size of the magnetic features, and the minimum length between the balls' centres, within each magnetic feature, is $1\px$ at any time. Nonetheless, in a practical case of instrumental data, the minimum ball spacing should account for the full-width-at-half-maximum (FWHM) of the point spread function (PSF) of the instrument. Should the pixel scale of the imager be smaller than the FWHM, the minimum spacing between the ball should be set to the nearest integer value. This will optimise the total number of balls, and thus the computing time.

Once the balls are positioned on the magnetic features, or more precisely, within the "magnetic holes", the polarity of each true feature is retrieved from the signed intensity of the original magnetogram, at the pixels mapped to the coordinates of each ball's centre. This polarity is stored, and is a constant associated with each ball. It is referred to as the initial "ball polarity". 
\\Next, a few integration steps, typically $10$ to $20$ depending on the size of the features, are performed between the first and second frame, so the balls have time to converge down into the local minima. For instance, in \Fig{balltrack_mag}, such a local minima in the first magnetogram (Magnetogram 1) would be found at $X^1_{\mathrm{final}}$.

Because a segmentation algorithm will be used on the tracked magnetic features (more on that later), it is not necessary to have several balls within the same feature. If several balls have converged to the same local minima, only one ball is kept. After this stage, it is still possible to have one large magnetic feature being tracked by several balls, if for instance the feature has several local minima. This is illustrated in \Fig{mballtrack_real_init} (bottom). Note that this significantly reduces the number of balls between the first (top) and the next frame (bottom), and consequently the computational time.

\subsection{Phase 3: main tracking phase  \label{sec:mphase3}}
After the initialisation, the next frames are loaded, and the balls track the local minima (the "magnetic holes") like they do within intergranular lanes with balltracking \citep{Potts04}. 
\\In the original magnetograms, the local minima of the rescaled data correspond, respectively, to the local maxima and minima of the signed intensity (positive and negative, respectively) of the magnetic flux density. At any time, the position of each ball is known, and each ball is labelled with a unique number, referred to as the "ball number". The positions can be plotted on-the-fly, so that one can check by eye the quality of the tracking. An example of the main tracking phase is visible in the  snapshots of \Fig{mballtrack_real}, with three ball numbers  identifying three distinct magnetic features that are being tracked.

When a magnetic feature is moving too rapidly, the balls do not have time to settle in the local minima. In this situation, at worst, the balls may be delayed by a few frames, and several integration steps between each frame are necessary to make sure that the balls do not get lost. This gives  them more time to "catch up" on the fastest features. Typically, for magnetograms taken at cadences of up to $3\minutes$, 10 to 20 intermediate integration steps of the equation of motion are used between each frame. This may or may not be the same number of intermediate steps used for the initialisation phase, these are indeed two independent "tuning" parameters that depend on the resolution of the data and the time sampling rate of the instrument. Typically, the number of integration steps is proportional to the former, and inversely-proportional to the latter. Indeed, at higher resolution, the features are relatively wider, and the balls must travel over a relatively greater number of pixels. At increasing time sampling rates, the features evolve less rapidly between two consecutive magnetograms, and so the balls need less time to catch up on the motions of the local minima. 

For large connected magnetic features, the shape of the magnetic features looks like an extended surface full of holes, where each hole can be filled with a ball. See for example the white magnetic patch in \Fig{mballtrack_real_init} (bottom) around the coordinates (60, 120) and the balls in the 3D view in \Fig{mballtrack3D} (bottom). This method is suitable for tracking clustered features that are made of several fragments, such as the ones in \Fig{mballtrack_real}.

At each tracking step, the polarity under the current position of the ball's centre is compared to the ball polarity. The tracking of a given ball ends as soon as the polarity of the current pixel is reversed with respect to the ball polarity. 
\\This strategy has several advantages. Indeed, to "see" a reversed polarity, a ball needs either to keep tracking down to the noise level until the current pixel polarity reverses (which is known by looking it up automatically in the original magnetograms), or it needs to encounter a magnetic feature of opposite polarity. This defines two conditions for the end of the tracking of a given ball. They are described below separately. 
 
\subsubsection*{Condition 1: tracking down to the noise level}
Tracking down to the noise level makes the algorithm use the true sensitivity of the instrument. Indeed, even if the initialisation step uses a threshold, the features are tracked until they cannot be detected by the instrument, i.e, to values below the threshold. Should we ever need to track the faintest magnetic features from the beginning, the threshold may simply be lowered down to the noise level, which has the only consequence of increasing the computing time (more balls will be added).  
When the ball is floating over random noise, the sign of the intensity in the original magnetograms eventually reverses. In this case the tracking of this ball ends. The stability of the ball within noisy data is set by the damping coefficient $\alpha$ (\Eq{ball_eq1}). Typical values are between 0.1 and 1, depending on the time sampling rate and spatial resolution. In term of damping time ( defined as $T_d = 1/\alpha$), this is equivalent to values between 1 and 10, in units of time interval between frames. The damping force is necessary not only in the presence of noise, in which case the tracking is, in fact, more resilient with damping than without, but it is also necessary for the general stability of the code, regardless of the noise. This is illustrated in more detail in \Section{sec:damping}.

\subsubsection*{Condition 2: no crossing of opposite polarity }
Another issue we had to solve is how to prevent the ball from crossing a feature with positive flux to a neighbouring one with negative flux. This would occur for example if the ball is in a local minimum, and if it has kept enough momentum to reach another close-by local minimum of opposite polarity at the next integration step. This can also occur, if the close-by local minimum moves quickly towards the ball, like for example in the case of the two footpoints of a loop being submerged. 

This problem can be thought of as a "numerical tunnel effect", in the sense that at one time a ball is in a magnetic hole, and at the next, it has crossed a barrier and lies within the second magnetic hole. The solution is as follows: because the true polarity of each pixel is known from the original magnetogram, the local minima are always associated with a polarity, which is compared to the ball polarity (see \S~\ref{sec:mphase2}) at the end of each integration step. Should a ball lie in a magnetic hole with a polarity opposite to the ball polarity, the tracking of this ball ends.\\

Conditions 1 and 2 are checked independently for each ball. If either of these two conditions is fulfilled for a given ball, its tracking ends, so that the lifetime of a ball corresponds to the lifetime of the magnetic feature tracked so far. This does not end the whole algorithm, which continues as long as other balls remain. Note also that in Phase 3, whenever there are overlapping balls, i.e, their centres are positioned on the same pixel, only the oldest ball is kept, which allows the right estimation of the features lifetime (more details on this in Phase 4).
\begin{figure}
\centering
	\includegraphics[width=1\columnwidth]{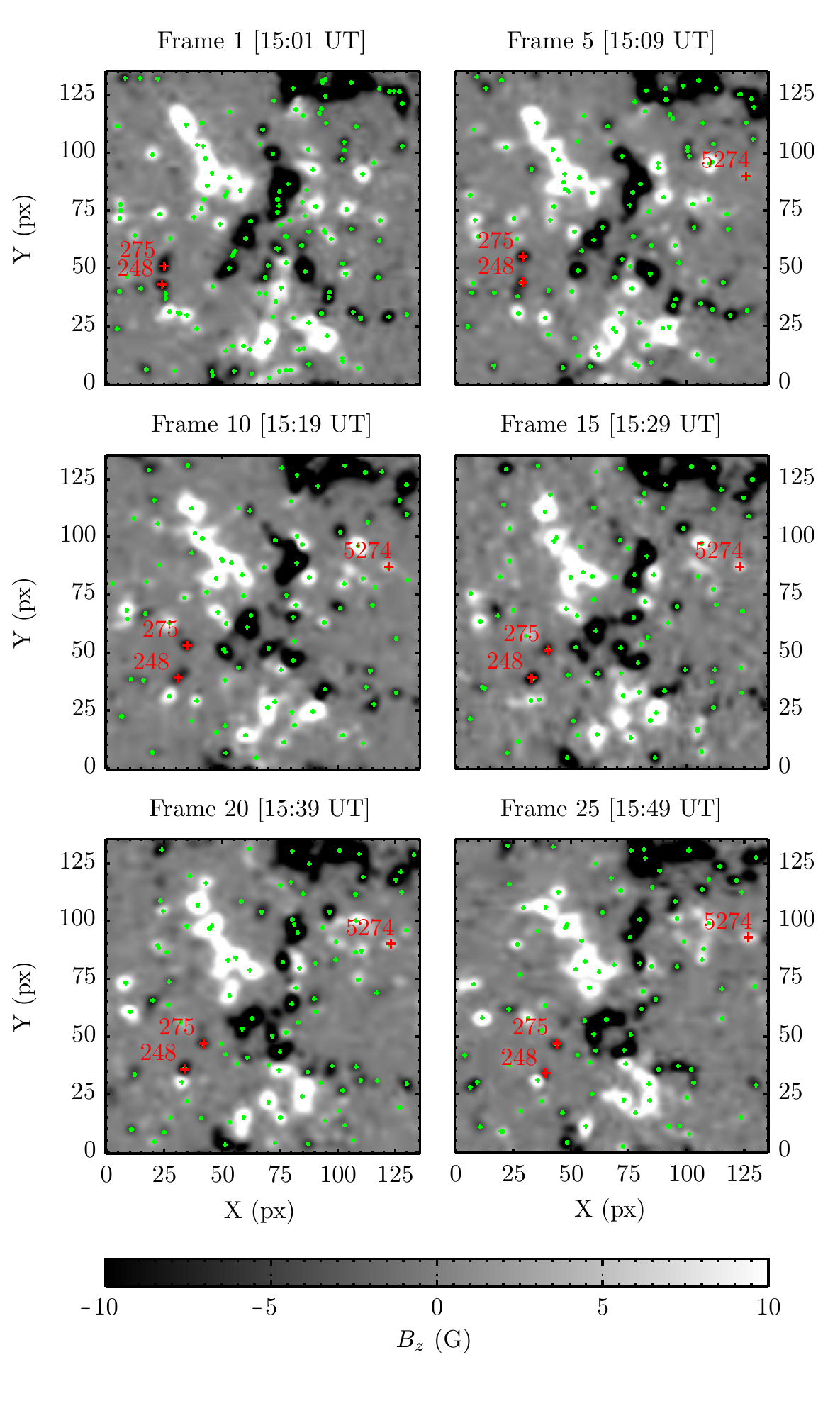}
	\caption{Magnetic balltracking of a small field of quiet Sun, at different time steps. The numbers bound to the crosses (red) are the labels that are unique to each ball (i.e. the ball number). \label{mballtrack_real}} 
\end{figure} 

Note also that the magnetic features may exist at very irregular places (see \Fig{mballtrack_real_init}, top), and consequently, the balls end up scattered over a rather irregular grid. Therefore, contrary to the original balltracking algorithm, the conversion of the velocity, initially calculated in a Lagrangian frame of reference, into a Eulerian frame of reference cannot happen. In other words, this method tracks and follows the individual motions of the magnetic features. It is not designed to output a "flow field", with the values of the velocity known at any given time at fixed positions, on a regular grid, like in, e.g., \citet{Demoulin2003} and \citet{Welsch2004}. This is simply because there are parts of the magnetograms where the magnetic flux is not detected, and thus, there is no velocity defined there. At best, Eulerian flow fields can only be derived locally, in regions where there are enough magnetic features that provide a more "reasonable" sampling.

\subsection{Phase 4: detection of emerging flux \label{sec:mphase4}}

Note that this phase is an optional module in the algorithm. When used, it runs simultaneously with Phase 3.
During this phase, the algorithm permanently scans for new pixels that would rise above a given "detection threshold", which must be at least greater than the noise level of the instrument. It may, or may not be equal to the threshold defined in Phase 2. When new pixels "rise" above this threshold, new balls are added on these areas so that emerging flux can be tracked. This phase needs another tunable parameter, referred to as the "detection spacing", which sets the minimum distance between the new balls and the ones already present. Any pixel whose intensity rises above the threshold must satisfy the condition of being at a distance larger than this parameter, in which case a new ball is put there. Note that this makes it also possible to follow large features whose size varies significantly over time. Indeed, the large features are the ones with many pixels above the detection threshold. If the distance between these pixels and those that have balls already, is greater than the detection spacing, new balls will eventually be put there too. Consequently, if a given feature, initially small and populated with one ball, grows over time (for instance, as a result of flux emergence), it may be populated by more than one ball. Note that the detection spacing also defines the resolution with which emerging flux is detected. The subsequent tracking of the new balls is exactly the same as the other balls. 

An example is given with the ball 5274 in \Fig{mballtrack_real} (starting from frame 5, near the upper right corner, seen as a small red cross). It is tracking emerging flux that, before frame 5, was below the initialisation threshold of $5\G$. When the magnetic feature emerges above $5\G$, this new ball locks onto it and tracks it until the last frame. This is allowed because the nearest balls tracking other features are at a distance greater than the detection spacing. So this phase makes the algorithm useful not only to track the flux visible from the start, but also to track the emerging flux, and little pieces of the largest features that may (or may not) fragment over time.

\begin{figure*}
	\centering
	\includegraphics[width=1\textwidth]{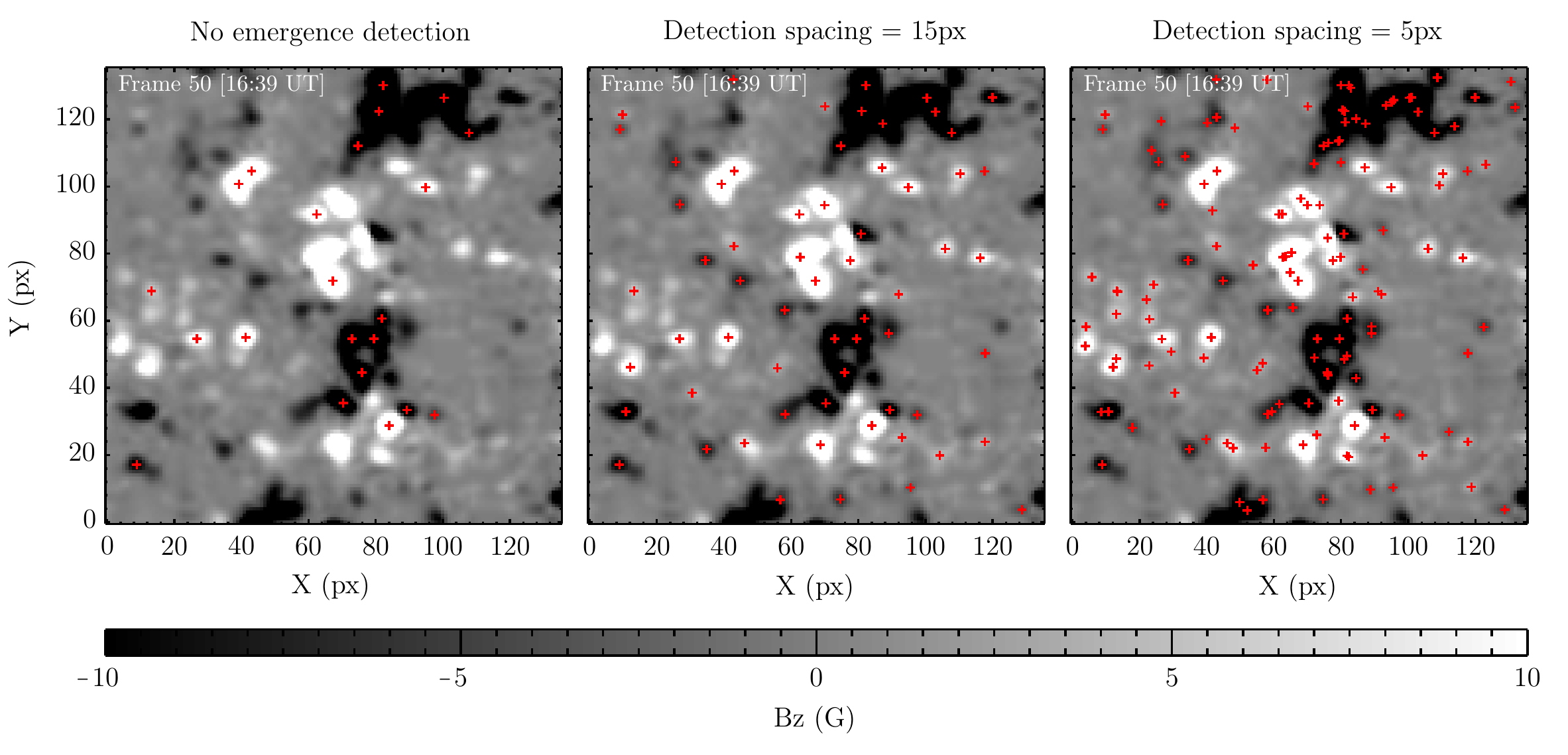} 
	\caption{Snapshot comparing the effects of the detection spacing, at a given detection threshold of $5\G$. The red crosses are plotted at the positions of the ball's centre. Left: no flux detection. Middle: loose detection spacing at $15\px$. Right: aggressive detection spacing at $5\px$. A movie showing the temporal evolution is available in the online edition.} 
	\label{snap1}
\end{figure*}

In the first movie attached to the online version of this paper, one can compare the effects of different values of the detection spacing (at a given detection threshold of $5\G$), and visualise Phases 3 and 4 of magnetic balltracking. A snapshot of this movie is shown in \Fig{snap1}. Three panels are shown. With these data, the pixel size is $0.2\arcs$ and the resolution is about $0.3\arcs \px^{-1}$ \citep{Chae2007}. In the left-hand panel, the detection of the emerging is deactivated. In the middle panel, the detection spacing is $15\px$, which is still a bit "loose", and one misses a few emerging features. Finally, in the right-hand panel, the detection spacing is set to $5\px$ which is a rather "aggressive" detection. The latter may be harder to follow because the finer the detection grid, the more balls are involved, and the more "crowded" these movies get. Nonetheless, it allows for more flux to be detected. A detection spacing of $10\px$ is typical. With a pixel size of $0.2\arcs$, which is about 6 to 7 seven times the size of the resolution element ($0.3\arcs \px^{-1}$). One can also see several balls gathering from the boundaries toward the centre of the largest features. This is due to new balls, added near the edges of the large features (Phase 4), converging toward the same local minima. Like in Phase 3, only the oldest ball is kept whenever, and wherever they overlap. This prevents "young" balls, i.e, the ones added during the flux detection, from replacing the older ones. Otherwise, we could not extract meaningful lifetimes during Phase 3.

Note that it is also possible to track emerging flux by simply reversing the timeline of the data series and recording submerging flux. In such a case, as far as the algorithm is concerned, flux removal is indistinguishable from flux emergence. Depending on the situation, this may be more suitable than using the module described in Phase 4. 

\subsection{Notes on the damping force \label{sec:damping}}
\begin{figure}
	\centering
	\includegraphics[width=1\columnwidth]{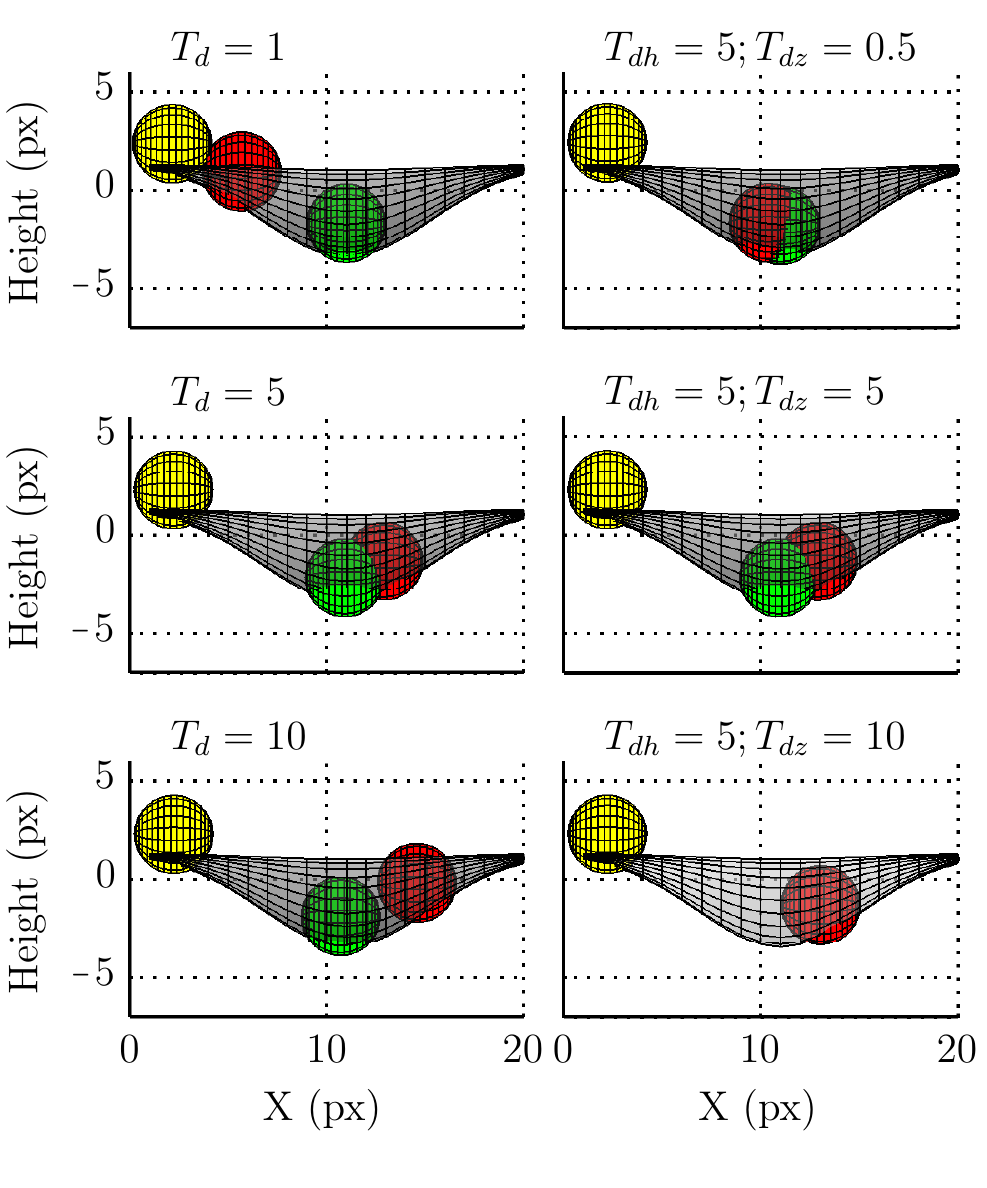}
	\caption{Tracking of the local minimum of a synthetic gaussian surface. The balls are projected on the x-axis and z-axis at three different numbers of integration steps. Step 1 (yellow), 15 (red), 100 (green). The left column uses the same damping time for the vertical and horizontal axis. The right column shows the tracking with a horizontal damping time ($\Tdh$) different from the vertical damping time ($\Tdz$). \label{damping3d}} 
\end{figure}

Without damping, the balls have too much inertia and "free-fall" toward the local minima. With too great a speed, they can move passed the local minimum. The damping coefficient prevents this "overshoot", and damps the subsequent oscillations of the positions when the balls "jiggle" around  the local minimum. This is illustrated in \Fig{damping3d} using a synthetic, rescaled gaussian surface, used as a magnetic feature that has a maximum strength of $50\G$, and a local minimum at $x = 11\px$. The FWHM of the original gaussian curve (i.e, before rescaling) is $\simm7\px$. On the left column, three values of damping times are used \mbox{($T_d$ = 1, 5, and 10)}. On the right column, we separated the damping times into a horizontal  ($\Tdh$) and a vertical ($\Tdz$) damping time that are associated respectively with the horizontal and vertical components of the damping force, whose usage are recommended in \citet{Potts04} (although in the case of tracking granules). The balls are plotted at the integration step 1, 15, and 100. Note the overshoot seen with the red ball at $T_d=5$ and $T_d=10$, as well as at $[\Tdh=5; \Tdz = 5]$ and $[\Tdh = 5; \Tdz = 10]$. The red ball has moved passed the local minimum and falls back to it at a later time (green). The overshoot is not seen with a shorter damping time in both directions $T_d=1$ and a shorter vertical damping time $[\Tdh=5; \Tdz = 0.5]$.

\begin{figure}
	\centering
	\includegraphics[width=1\columnwidth]{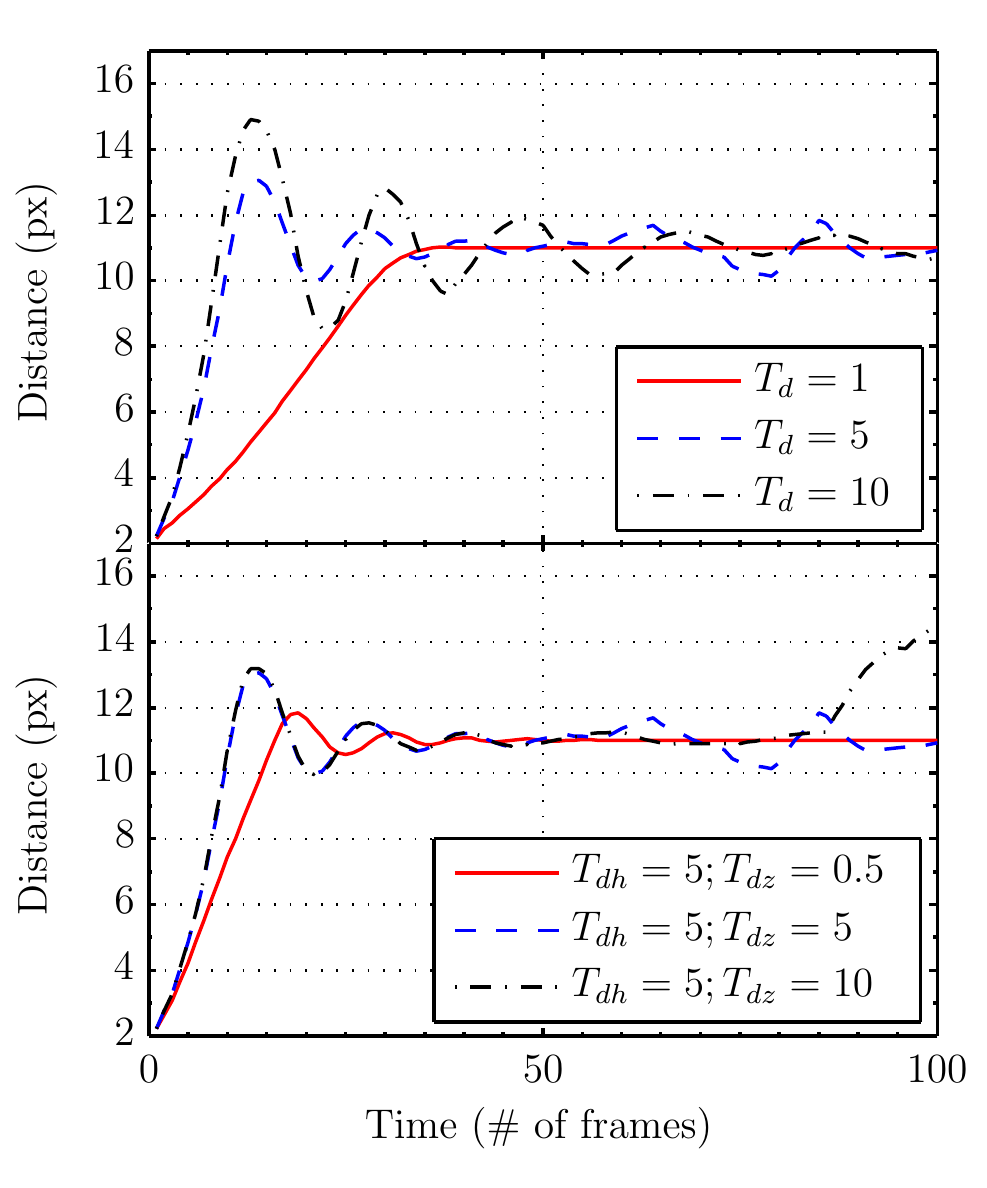}
	\caption{Coordinate on the x-axis of the ball's centre during the tracking of \Fig{damping3d}. The top and bottom panel correspond to the motions of the ball in the left and right column of \Fig{damping3d}. \label{damping1d}} 
\end{figure}

The x-coordinates of the ball in each case are plotted in \Fig{damping1d}. The top and bottom panel (resp.) correspond to the displacements the balls in the left and right column (resp.) in \Fig{damping3d}. The overshoot corresponds to $x>11$. Note how the oscillations are reduced more efficiently with a decreasing damping time. At $T_d=1$, there is indeed no overshoot, and the ball settles in the local minimum after 30 steps. Yet it may be more efficient to use $[\Tdh = 5; \Tdz = 0.5]$, which induces a small overshoot, while within 25 steps the ball is less than $1\px$ from the local minimum. $[\Tdh = 5; \Tdz=10]$ is a limit case where the damping force on the vertical axis is too small such that the tracking does not converge (the ball is not visible in \Fig{damping3d} due to its z-coordinate that is off the grid). If efficiency is not an issue, the use of a unique damping time $T_d=1$ is sufficient, as long as the convergence criterium is fulfilled (see later in \Section{sec:rescaling}). However, for an optimum stability on the vertical axis, and based on many experiences (not shown here), we recommend the use of $\Tdz$ within [0.5;1], and $\Tdh$ within [1;10], which may vary according to the data characteristics: resolution, pixel size, cadence, noise level, and sizes of the features.

\section{Motivations for the rescaling method and convergence criterium \label{sec:rescaling}}

The choice of the rescaling method of the magnetograms is driven by two objectives: (i) a successful tracking of the features, and (ii) a minimisation of the number of integration steps that are necessary to achieve (i). 
In this section we demonstrate the advantages of the choice of rescaling defined in \Section{sec:mphase1}.

\subsection{Formal definition of the rescaling methods}

Our "generic" rescaling method formalises as follows: 
\begin{align}
 B^*_z(x,y) &= \mathrm{max} (\,|B_z(x,y)|^\gamma\,)-|B_z(x,y)|^\gamma \label{eq:Bz}\\
 B^*_{zn}(x,y) &=\frac{\displaystyle B^*_z(x,y)-<B^*_z(x,y)>}{\displaystyle \beta} \label{eq:meanBz} 
\end{align}
in which we have introduced a normalisation factor $\beta$, and $\gamma$: linear rescaling is equivalent to $\gamma = 1$ and non-linear rescaling equivalent to $\gamma \neq$ 0 and 1. Then we compare the following three scaling methods:
\begin{enumerate}
\item $[\gamma = 1 ; \beta = 1]$: no rescaling is actually performed.
\item $[\gamma = 1 ; \beta = \sigma_{B^*_z}]$: linear rescaling.
\item $[\gamma = 0.5 ; \beta = \sigma_{B^*_z}]$: non-linear rescaling.
\end{enumerate}
Method 3 is the method defined in \Section{sec:mphase1}. $\sigma_{B^*_z}$ was originally defined as the standard deviation of $B^*_z(x,y)$, it implicitly depends on $\gamma$ (\Eq{eq:Bz}). Note that subtracting by the mean value in \Eq{eq:meanBz} only changes the zero-point of the data surface and does not affect the results. It is used here to conveniently define the mean value of the data as the origin of the vertical axis.

\subsection{Convergence criterium \label{convergence}}

\begin{figure}
	\centering
	\includegraphics[width=1\columnwidth]{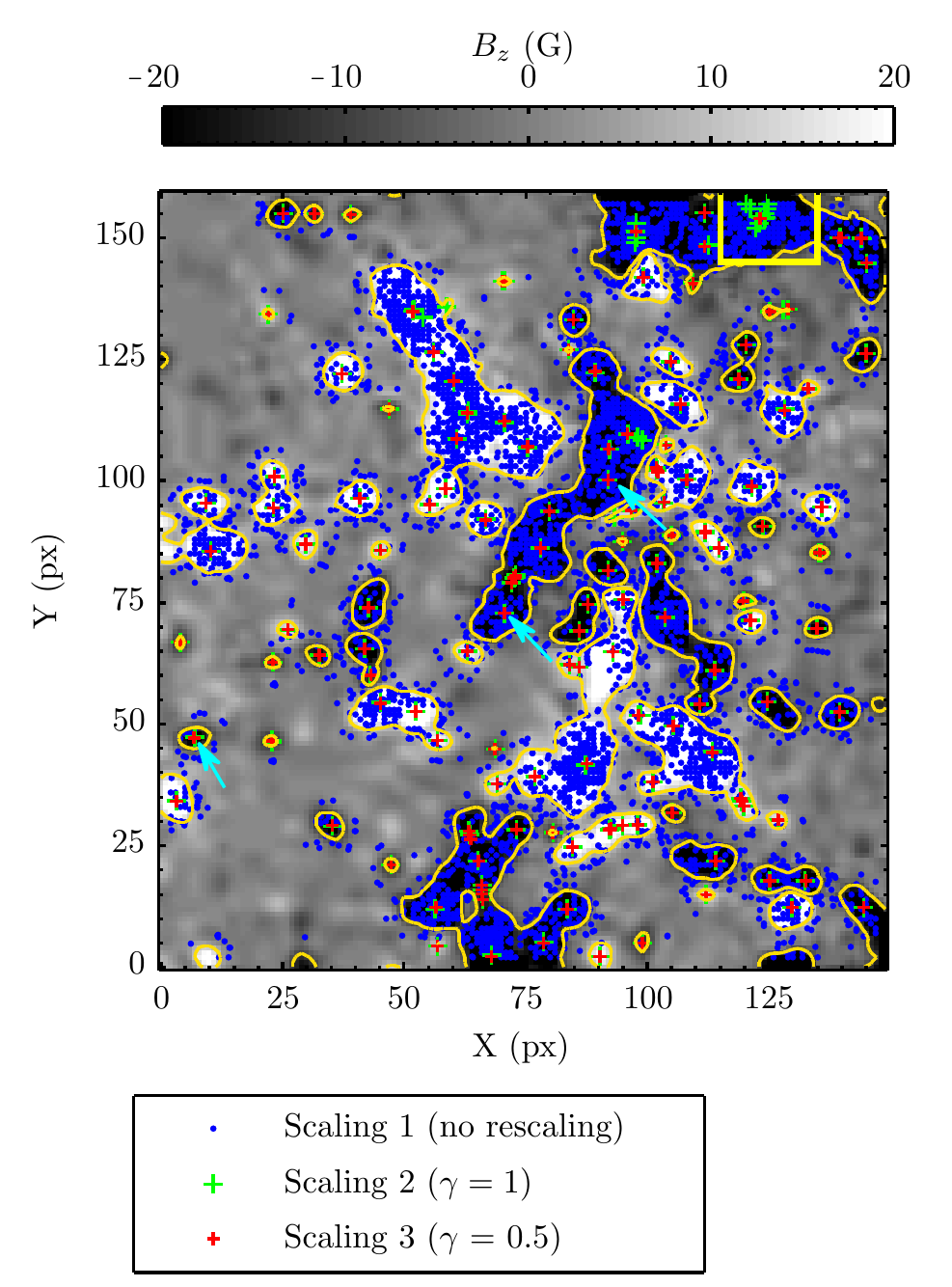}
	\caption{Final positions of the balls' centre, after 100 integration steps, for the three scaling methods. The original positions are the ones shown in \Fig{mballtrack_real_init} (top, red crosses). \label{overviewFinalPos}} 
\end{figure}

The positions of the balls at the end of 100 integration steps are shown in \Fig{overviewFinalPos} for the three scaling methods 1, 2, 3, respectively,  in blue dots, green and red crosses, respectively. The yellow contours are set at the initialisation threshold of $10\G$. The initial positions are the ones plotted in \Fig{mballtrack_real_init} (top). In the current example, a successful tracking has the necessary (but not sufficient) condition that all the final positions are within the yellow contours in \Fig{overviewFinalPos}. Yet the blue dots end up outside the contours (i.e, outside the magnetic feature they are meant to track), whereas the others are all within the contours, which proves that the scaling method 1 does not result in a successful tracking, and that Methods 2 and 3 are more appropriate.

\begin{figure}
	\centering
	\includegraphics[width=1\columnwidth]{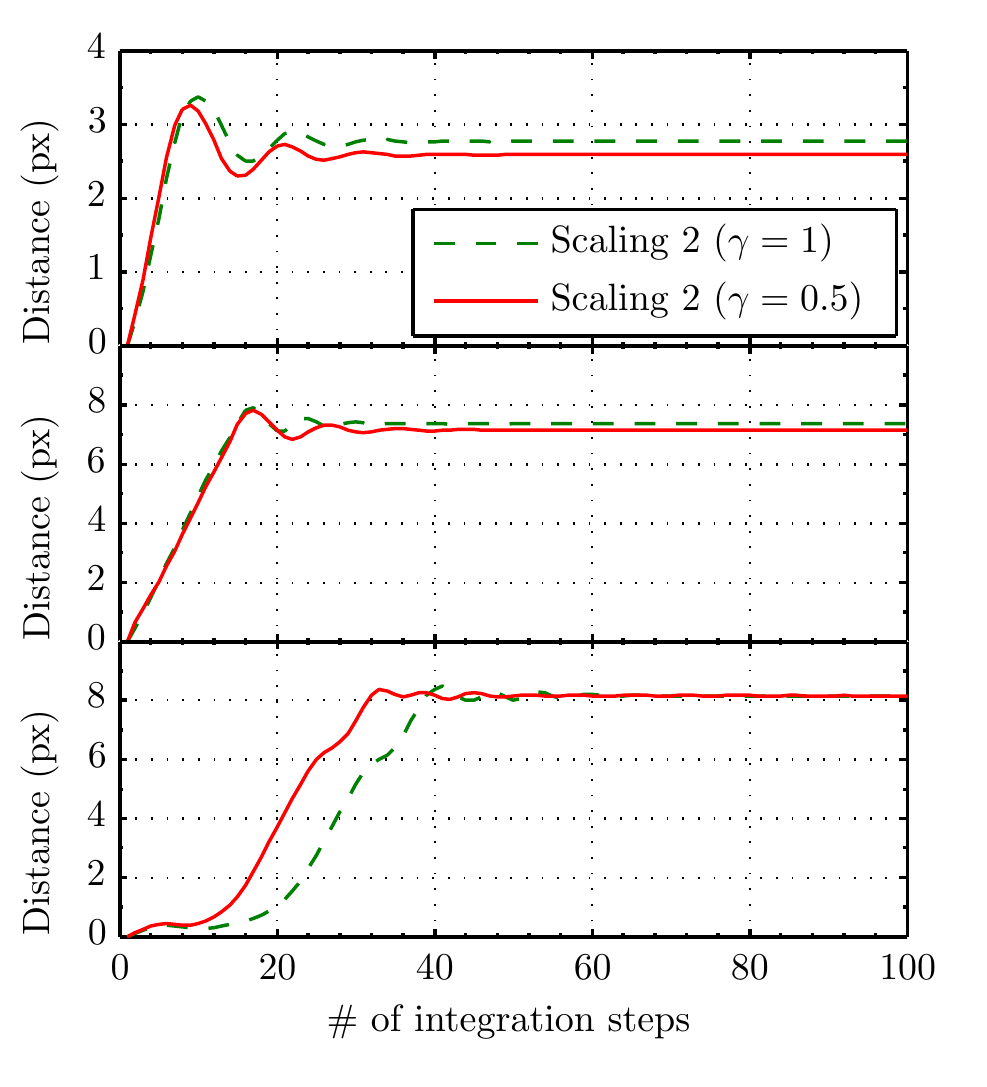}
	\caption{Distance from the initial position, for the three balls (resp. top to bottom) pointed to by the cyan arrows in \Fig{overviewFinalPos} (resp. from left to right). \label{FinalPos3}} 
\end{figure}

In \Fig{FinalPos3} we have plotted three examples of the traveled distance (in pixels) of three different balls' centres, using the linear (green dotted line) and the non-linear (red continuous line) scaling. The chosen balls, respectively from top to bottom, correspond to the ones pointed to by the cyan arrows in \Fig{overviewFinalPos}, respectively from left to right. The oscillations are due to the balls overshooting to either side of the local minima before they can finally settle. The oscillations are damped within one to two times the horizontal damping time (here it is set to 4).

\begin{figure}
	\centering
	\includegraphics[width=1\columnwidth]{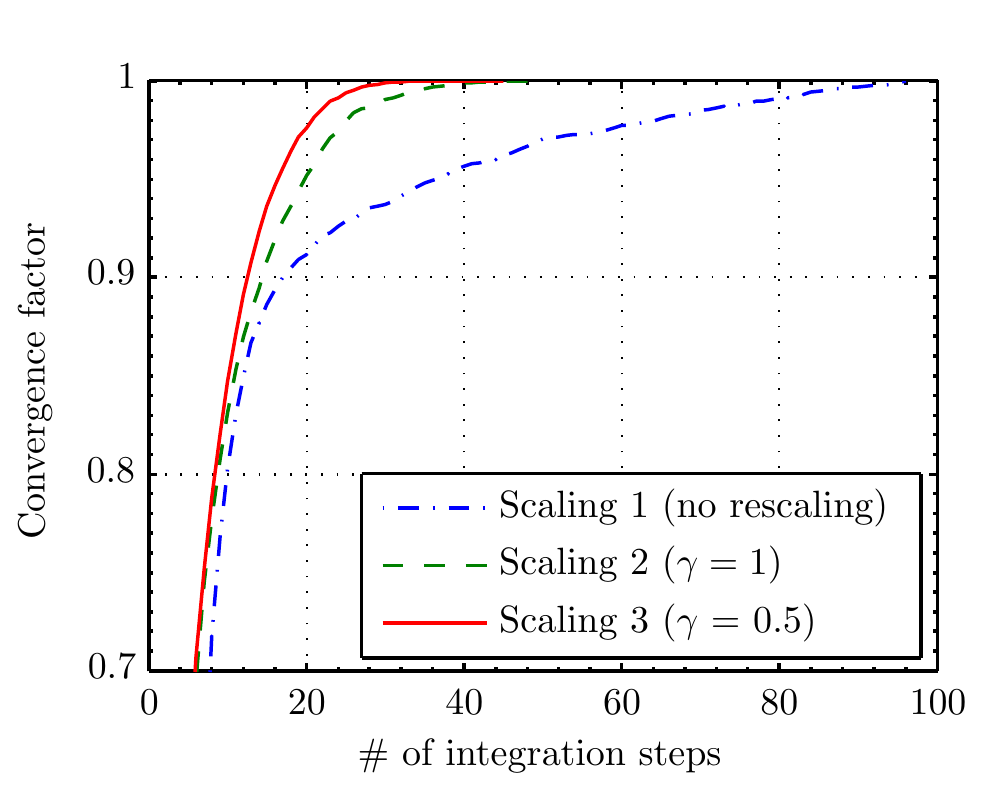}
	\caption{Convergence factor against the number of integration steps, for the three scaling methods. \label{ConvFactor}} 
\end{figure}

The time that the balls take to reach their final position defines a convergence factor represented in \Fig{ConvFactor}. The "final position" is set as the position reached by the ball after a sufficiently high number, here set to 100. The convergence factor is then defined as the ratio of the number of balls that have reached their final position at a given integration step, to the total number of balls. When the convergence factor reaches 1, the convergence criterium is satisfied, and this sets the number of initialisation steps. The scaling method 1 converges after about 95 integration steps, the other two methods converge in less than 40 integration steps, with the non-linear rescaling (red continuous line) converging more rapidly by a few integration steps. With the scaling methods 2 and 3, 90\% of the balls have converged within 10 to 20 steps, and all the balls have converged within 40 steps (the convergence factor equals 1). 20 initial steps is typical but a "trial-run" like the one here must be done in each case study to determine the optimum value for a given data set. Here, the optimum value would be 40.

\subsection{Comparisons between the rescaling methods 2 and 3}

\begin{figure}
	\centering
	\includegraphics[width=1\columnwidth]{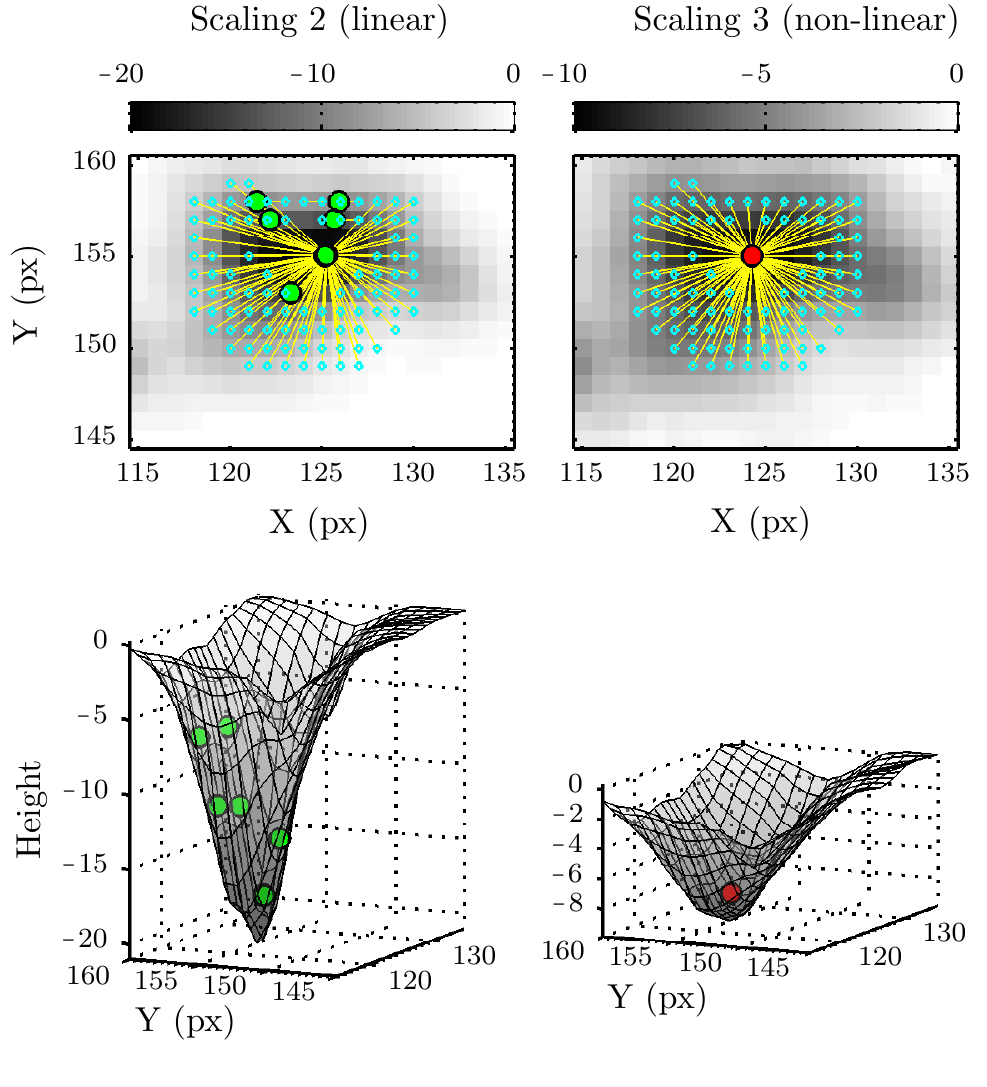}
	\caption{Close-up in the region of the yellow rectangle in \Fig{overviewFinalPos}. Left: linear rescaling (Method 2). Right: non-linear rescaling (Method 3). The small cyan dots are the initial positions. The coloured circles are the final positions. The yellow lines are mapping the final position to the corresponding initial positions. They are not the actual trajectories. Bottom: 3D data surface viewed from the left Y-axis. \label{rescale4Panels}} 
\end{figure}

At some places, like in the yellow rectangle (top right) of  \Fig{overviewFinalPos}, the green crosses are not overlaid by the red ones, and instead, a few of them ended up at different positions within a few pixels. This first suggests that whereas the linear rescaling (Method 2) has converged according to \Fig{ConvFactor}, the balls, in fact, have not converged in a local minimum, contrary to Method 3 where the red crosses suggest that the balls have unambiguously settled in the same local minimum. \Fig{rescale4Panels} is a close-up in the yellow rectangle of \Fig{overviewFinalPos} viewed from the top (top panels) and from the side (along the left Y-axis) of the 3D rescaled data surface (bottom panels). The coloured circles are plotted at the final positions of the balls' centre. One sees that there is indeed only one local minimum in this area, but the balls in Method 2 (left) did not settle in it. Some of them are standing still in the middle of the steep slope (bottom left), which is not what one might expect in a real "natural" situation. This is a typical example of a linear rescaling being less appropriate than a non-linear rescaling with $\gamma = 0.5$ (right). With a linear rescaling, the strongest magnetic features (here above $100\G$) still have a very steep slope (bottom left), while the slope is more gradual with a non-linear rescaling (bottom right). 

\begin{figure}
	\centering
	\includegraphics[width=1\columnwidth]{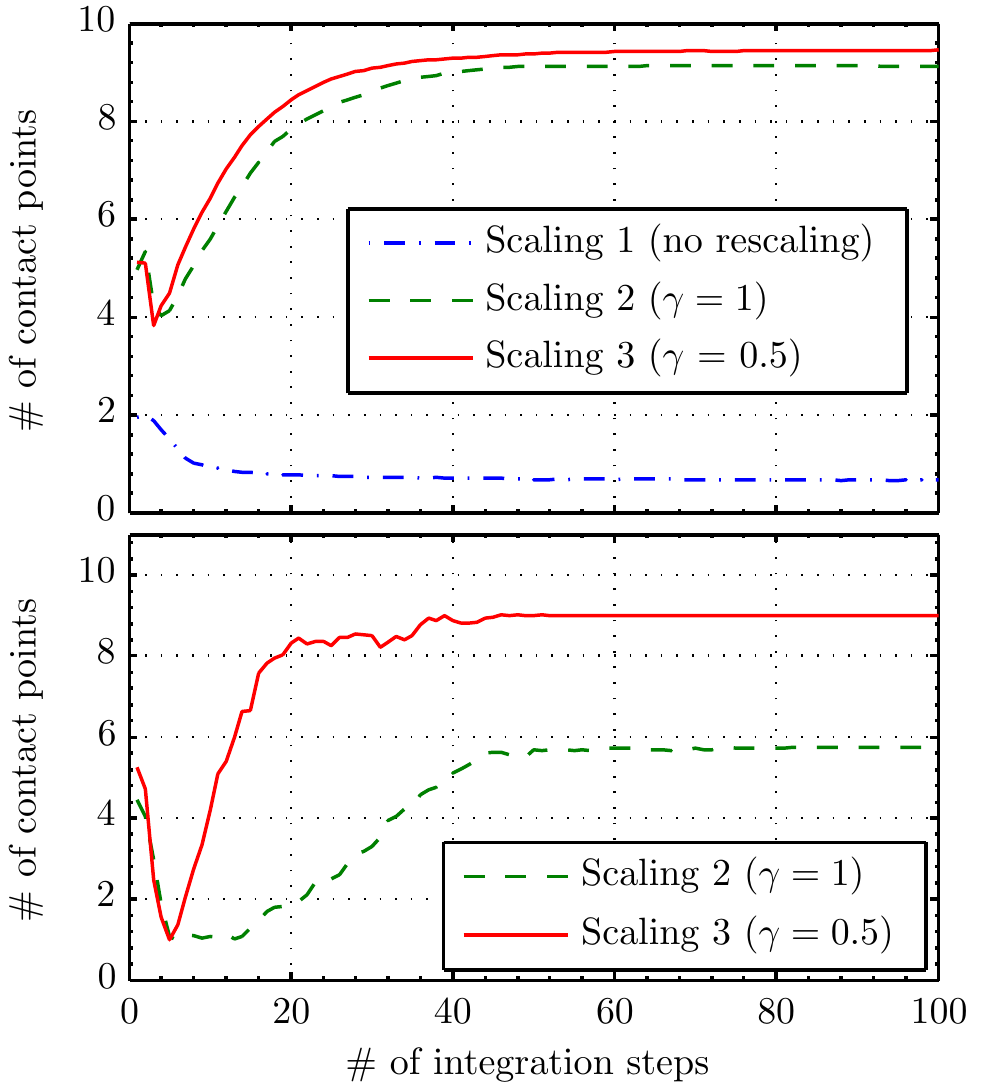}
	\caption{Average number of contact points per ball, against the number of integration steps. Top: average over the FOV of \Fig{overviewFinalPos} for the three scaling methods. Bottom: average over the balls in \Fig{rescale4Panels} for the linear and non-linear scaling methods. \label{ContactPts}} 
\end{figure}

The reason for the very steep slope resulting in an inconsistent tracking is related to the number of data points that are actually in contact with the balls, and therefore, contributing to the total force $\mathbf{f_i}$ (see also \Eq{ball_eq1} and the blue squares in \Fig{balltrack_mag}). Indeed, for maximum efficiency, the 3D grid defined by the balls, and which samples the data surface, uses regularly spaced values with a grid size of $1\px$. For a data point to be considered by the algorithm, it needs to be located within one ball radius from the ball's centre. The data at these ball grid points are interpolated. Note that nearest neighbour or linear interpolation make no difference on the final positions (the former is preferred for efficiency). At a given ball radius, the more of these "contact points", the better, as more contact points means that more of the magnetic feature actual topology is accounted for, and thus the more coherent the tracking is. 
The number of contact points involved in the tracking in the three different scaling cases is shown in \Fig{ContactPts}. In the top panel, we use the  number of contact points averaged over all the balls involved in \Fig{overviewFinalPos}. 
The top panel again shows that in the case of no rescaling (blue dashed line), there is on average merely 1 contact point, which explains why the scaling method 1 fails. Methods 2 and 3 differ only slightly on average, by only 1 contact point. Locally however (bottom panel), in the cases corresponding to the close-up in \Fig{rescale4Panels}, this difference is more significant. Before the convergence criteria is reached (i.e, before the integration step 40), the number of contact points varies from 1 to less than 6 with a linear rescaling, whereas it varies from 1 to more than 8 with a non-linear rescaling. Furthermore, with the latter method, this number increases more rapidly: it equals 8 while it is only 2 with the linear rescaling, which can only result in an insufficient force to push the balls down the slope. Put simply, with too few contact points, the balls can stall. 

Thus, Method 3 has two advantages over Method 2: a better handling of the steeper slope for the magnetic features with field strength of the order of $10^2\G$, and a relatively higher efficiency (a few integration steps less are needed). The tracking of faint magnetic features is not affected by these effects. Visual inspections show no difference in the final positions between Methods 2 and 3 when tracking the smallest and weakest features close to the initialisation threshold. In addition, no case where Method 3 is less accurate than Method 2 were found, and therefore, the non-linear rescaling is the most appropriate choice among the three tested methods. Note that other values of $\gamma$ within $]0;1]$, but relatively close to 0.5, are possible, and they may be tuned to different values for optimisation purposes with quiet Sun features, with negligible changes on the final positions.

\section{Effects of the balls radius \label{Radius}}

\begin{figure}
	\centering
	\includegraphics[width=1\columnwidth]{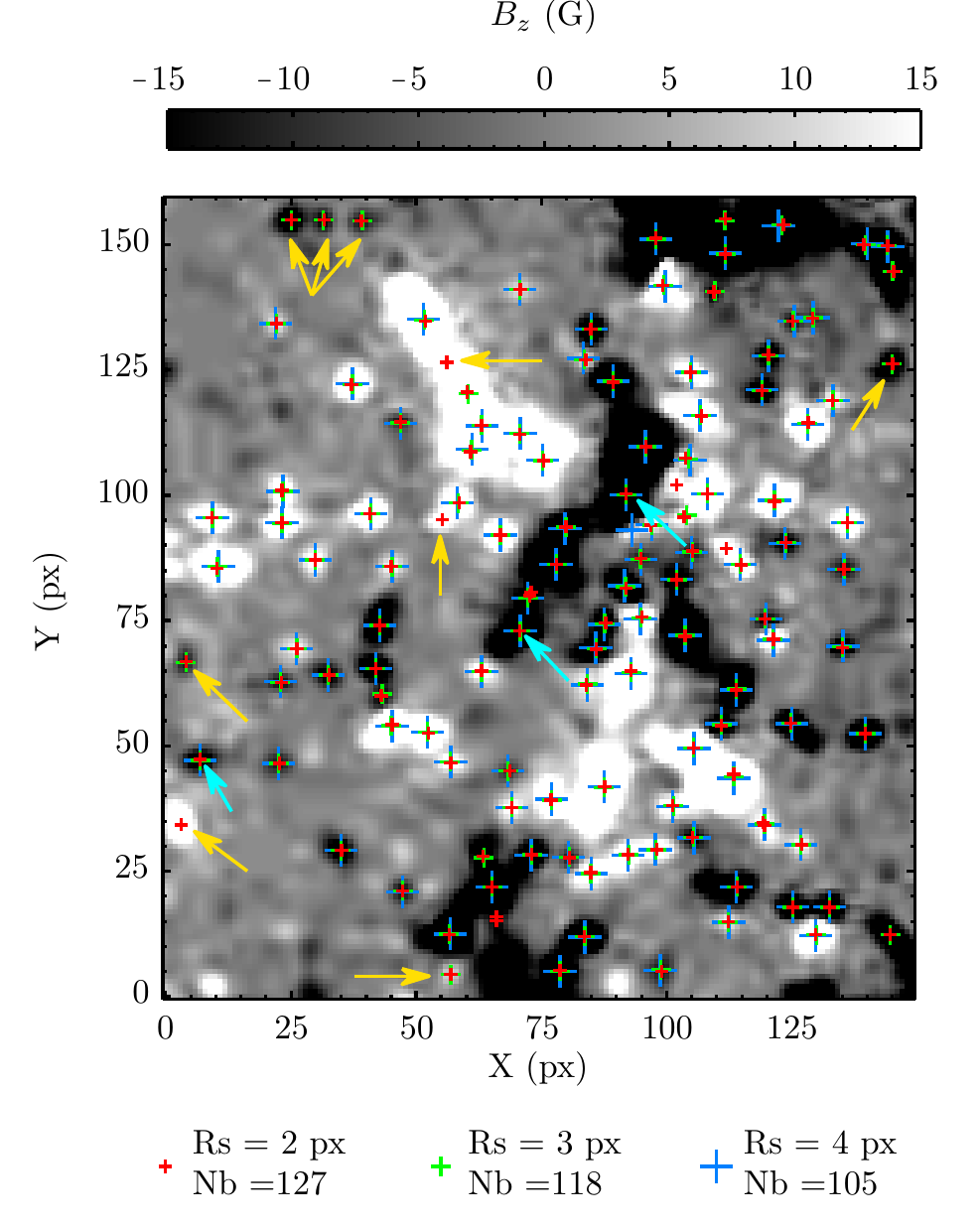}
	\caption{Final positions using three different ball radii: $Rs=2\px$ (small red crosses), $Rs=3\px$ (medium-sized green crosses), and $Rs=4\px$ (large blue crosses). "Nb" is the number of balls visible in the figure for each ball radius. The yellow arrows point at local minima in which only the balls with a 2-px-radius converged. \label{RadiiOverview}} 
\end{figure}

The appropriate choice of the ball radius relates to the size of the magnetic features that are to be tracked. The sphere radius must be greater than 1~px, regardless of how small the features are. At 1~px the number of contact points is too small to result in a coherent tracking, which stays dominated by high-frequency noise. \Fig{RadiiOverview} shows the final positions of a tracking using three different radii (Rs), represented by coloured crosses of increasing sizes: Rs=2~px, Rs=3~px, and Rs=4~px. "Nb" is the number of different local minima in which the balls have converged. It is here equal to the number of crosses of a given colour that are visible in the figure. Nb is respectively equal to 124, 117, and 103. So the number of different local minima in which the balls managed to settle decreases when the radius increases. The "missing" balls at Rs=3~px, and Rs=4~px have either "fallen off" the edges and/or passed "through" the smaller ones located near bigger local minima in which more of the bigger balls are pulled in more efficiently: the yellow arrows show examples of local minima resolved by the use of a 2~px-radius, but that are not resolved by the use of Rs=4~px and/or Rs=3~px. There is one place where  a ball with only the 4-px-radius converges (near the coordinates (100, 100)) and it is, in fact, a "false positive" (i.e, not an actual local minimum): the lack of resolution made the bigger grid (due to the greater radius) cover pixels on the neighbouring feature (black patch on the left), which averages the motion of the tracked white patch and of this nearby black patch. On the other hand, we do not find any "false positive" with the red crosses (Rs=2~px), which makes it, here, the best choice.

Note that these test data are from NFI/SOT (Hinode). The data were binned onboard to a pixel size of $\simm0.2\arcs$ while the spatial resolution of NFI magnetograms is $0.3\arcs \px^{-1}$ \citep{Chae2007}, so a radius of 2~px or 3~px would be appropriate. Nonetheless, for a given initial ball spacing (defined as the minimum space between the balls at the initialisation phase), using a greater radius reduces the resolution of the tracking, and also increases the computing time and memory usage as a bigger ball encompasses more grid points.

\begin{figure}
	\centering
	\includegraphics[width=1\columnwidth]{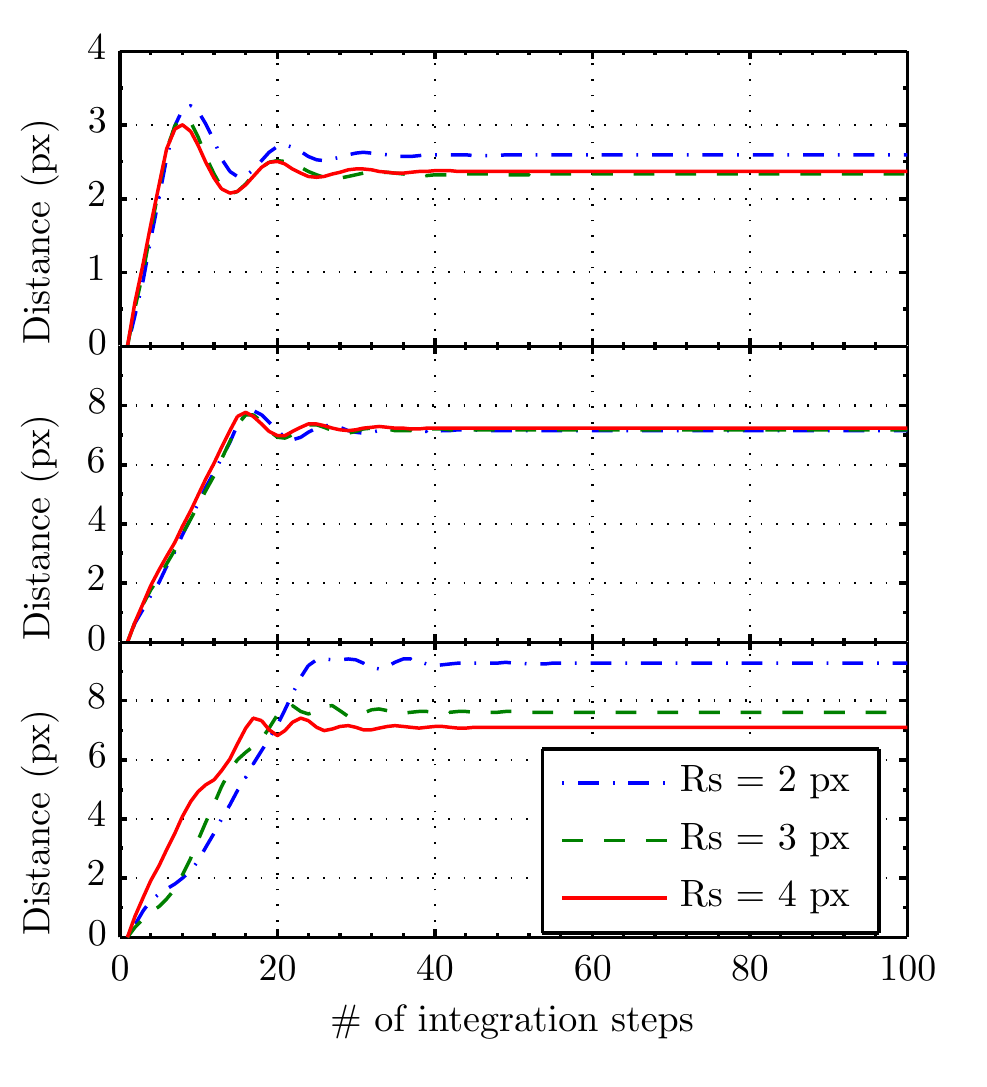}
	\caption{Travelled distances in three cases, with three different ball radii: $Rs=2\px$,$Rs=3\px$, and $Rs=4\px$. From top to bottom (resp.), the balls ending up at the location shown by the cyan arrows from left to right (resp.) in \label{FinalPos3Radius}.} 
\end{figure}

\Fig{FinalPos3Radius} is a comparison between the horizontal distances travelled by the balls at three locations chosen at random (from top to bottom, respectively) pointed to by the cyan arrows in \Fig{RadiiOverview} (from left to right, respectively). In each case, the three radii are tested. The distance has its origin at the initial position of the balls, and in each case, the three balls have the same initial position. \Fig{FinalPos3Radius} shows there is less than one pixel of discrepancy in the final position between Rs=2~px and Rs=3~px. This discrepancy is greater with Rs=4~px, about 2~px according to the bottom panel. Looking at other cases (not shown here) gives the same conclusion that  Rs=2~px and Rs=3~px have the least discrepancy of less than one pixel, and that Rs=4~px is less accurate. Note that because the size of the features, in pixels, also depends on the plate scale and on the resolution of the instrument, we cannot give ideal values that work in all situations. Such tests on subsets of large data with different radii are necessary to optimise this choice. In addition, one could combine the results of different radii for multi-scale analyses. The small radii would be dedicated to the smaller features, while the greater radii would be tracking the broader patches. 

The main output of magnetic balltracking is the series of final positions of the balls, for the purpose of tracking the time-dependent displacements of the local extrema of the flux. These positions are then used in a segmentation algorithm to integrate the flux over the area of the tracked features. This is detailed in the next section. 

\section{Segmentation of the magnetic features \label{sec:regiongrow}}

As mentioned earlier, the magnetic features can be very clustered in the quiet Sun. Then quantifying the evolution of each bit of each magnetic structure, which can be near the limit of the instrumental resolution, is quite challenging. Magnetic balltracking facilitates this procedure. As explained in \Section{sec:phases}, this technique tracks the time-dependent positions of the local extrema of individual magnetic features. Thus the next step in describing their evolution is to integrate the magnetic flux of these features. An easy way to do this is by applying a "region extraction" algorithm. The technique is also known as "region growing", and has many names and variations that depend on the scientific field in which this segmentation technique is applied. It is one of the basic algorithms detailed in textbooks of digital image processing \citep[see for example][Chapter 10, \S~10.4]{Gonzalez2008digital}. 

\begin{figure}
	\centering
	\includegraphics[width=1\columnwidth]{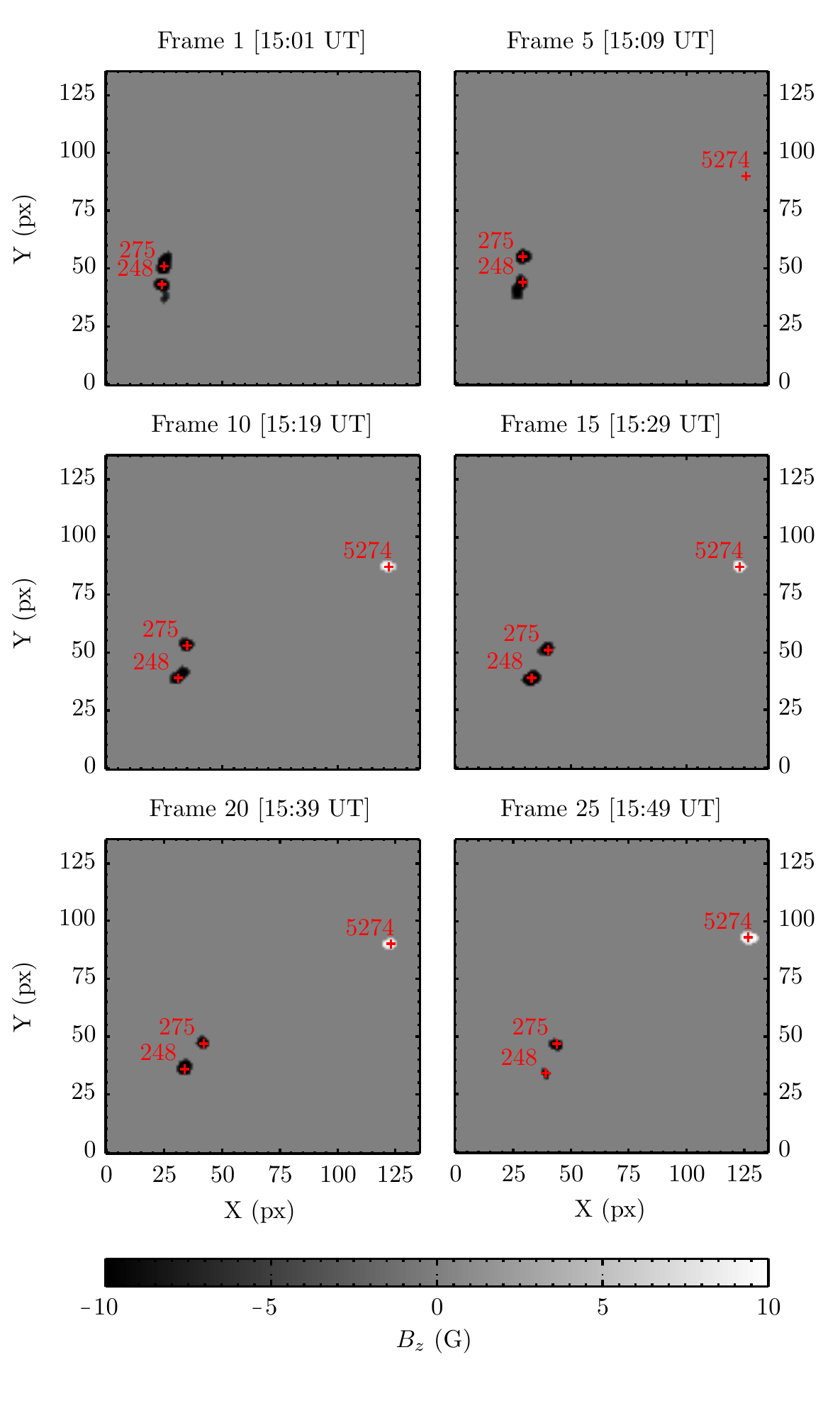}
	\caption{Region of the magnetograms extracted using the region-growing algorithm onto the balltracked seed-points of \Fig{mballtrack_real}. \label{mballtrack_real_mask}}
\end{figure}

The "region growing" used here, consists in extracting the magnetic features that have been tracked, from the rest of the magnetograms, so we can easily integrate the intensity of the flux over the extracted area. We use the tracked positions (the final positions of the balls in each frame) as so-called "seed points". For each position (or each "seed point"), the difference between the intensity of the neighboring pixels and the seed pixel is compared to a given threshold. If the comparison is true (in the logical sense), it is added to a list of pixels connected with each other, including the seed, which "grows" a region and thus segments the magnetic feature from the rest of the magnetogram. If the comparison is false, the pixel is not added to the list. The region stops growing when there are no more connected pixels. The output of the region-growing algorithm  is a binary mask: an array of logical values, co-spatial with the magnetograms, where the pixels in the grown region are set to 1, and the other pixels are set to 0. These masks can directly be used to extract the features from the magnetogram in order to integrate the flux of the tracked features. Such extracted features are visible in \Fig{mballtrack_real_mask}, which used the tracked positions (i.e, the seeds) previously illustrated in \Fig{mballtrack_real} as the input of this segmentation.

As mentioned in the previous section, several balls may track the same wide magnetic features, with as many balls as there are local extrema in it. As the position of these balls are used as seed points, they will ultimately extract the same connected pixels, and output identical masks. Then we have to get rid of the duplicates, which we do by using a logical "or" (equivalent to a logical "union") between all the extracted masks. If the same masks of connected pixels are output for different balls, the logical "or" reduces them to one unique mask before integrating the flux. This makes sure that, when looping over the extracted regions to integrate their flux, we do not integrate it over the same region more than once.

Finally, each magnetogram is integrated over the extracted areas, which gives the flux carried by the tracked feature in each single frame. Repeating this for all the magnetograms provides the time-dependent flux of all the tracked features, including the emerging one if any has been detected during Phase 4. The ball numbers are used as the labels of the magnetic features, which makes it easy to select them individually. This way we can choose which one to extract with the region-growing algorithm, but it is also possible to simply take all of the tracked features for more global statistical analyses.

One limitation of this method is that it cannot grow a region using too low thresholds, otherwise, close but probably non-connected regions may be added to the list of "good" pixels. This leads to a wrong segmentation, and ultimately a biased estimation of the flux. This could probably be solved using a more sophisticated segmentation algorithm. For consistency, we use thresholds close, but not necessarily equal to the thresholds used in the previous phases of the magnetic balltracking (defined in Phase 2). They are between, typically, $5\G$ and $15\G$. We found these values by "trial and error" and they turned out to be optimal in the very clustered flux of the quiet Sun. A direct unfortunate consequence is that no flux is integrated below these thresholds. Nonetheless, the actual value of the thresholds that we have used so far depend more on the instrumental detection limits rather than being an intrinsic limitation of the algorithm.

\subsection*{Difference with other algorithms}
To identify the magnetic features, the four codes compared in \citet{DeForest2007} use a similarity metric in the (x,y,t) space (2 spatial dimensions and 1 time dimension) based on logical comparisons between the shape and position of the segmented areas. This is done after the segmentation in the region of interest. The features are then identified and labeled as magnetic entities at the end of this process. These are then the "tracked" magnetic features. The motion of the features may or may not be derived using centre of gravity of the extracted area. 

In magnetic balltracking, the magnetic features are already labelled at the first phase, and in Phase 4 for the emerging ones, by the ball numbers that settle within their local maxima. Furthermore, the tracking is done before the discrimination of the magnetic fragments, using the balltracking paradigm. The latter occurs as a post-processing of the magnetic balltracking. Here we have used region growing as the post-processing method to build the masks that extract the magnetic fragment from the images, which is similar to the so-called "clumping" used in MCAT. Regardless, the magnetic balltracking can be combined with other methods of segmentation, as long as they can use the balls as seed points. Within magnetic features large enough to have more than one local extrema, and thus more than one ball tracking the whole feature, the duplicated masks associated with each of the ball (used here as seed points) are unified into one single mask. Further discrimination within these large features cannot be achieved with this method, and depending on the science goals, one may revert to using a more discriminative method like the so-called "downhill" method in YAFTA and SWAMIS. 

In the next section, we demonstrate the capabilities of magnetic balltracking in a case study of flux emergence using Michelson Doppler Image (MDI) data and co-spatial X-ray images from the X-Ray Telescope (XRT) on Hinode. 

\section{Magnetic balltracking on flux emergence \label{sec:flux_emergence}}

\subsection{Observation of flux emergence}

In what follows, all times are given in Universal Time (UT).
The observations were made on September $26^{\mathrm{th}}$, 2008, and consist of high-resolution, $1\minutes$-cadence continuum images and magnetograms from MDI \citep{Setal95}, and co-spatial images from XRT \citep{Golub07} in soft X-ray (pixel size of $1\arcs$), at $\simm30\s$-cadence. The different time series last 4.5 hours, between 15:00 until $\simm$19:30. The MDI magnetograms were rigidly de-rotated using the local latitude at the centre of the field-of-view (FOV), which is consistent with the preprocessing recommendation in \citet[\S~5.1]{DeForest2007}. The time series of XRT images were calibrated and registered using the routines related to the XRT instrument in Solarsoft (xrt\_prep.pro, xrt\_jitter.pro). The co-alignment of the XRT images onto the MDI magnetograms was done using co-temporal XRT images and Extreme-ultraviolet Imaging Telescope (EIT) images \citep{Delaboudiniere1995} whose frame-of-reference is put into alignment with the MDI frame-of-reference using the header information.  Accuracy of the latter was checked using limb fitting of the full solar disk. The XRT and EIT frame used for the co-alignment were taken near 19:00. We estimate the XRT and MDI frames to be co-spatial within less than $1\arcs$ ($<1\Mm$) . 

\begin{figure}
\centering
	\includegraphics[width=1\columnwidth]{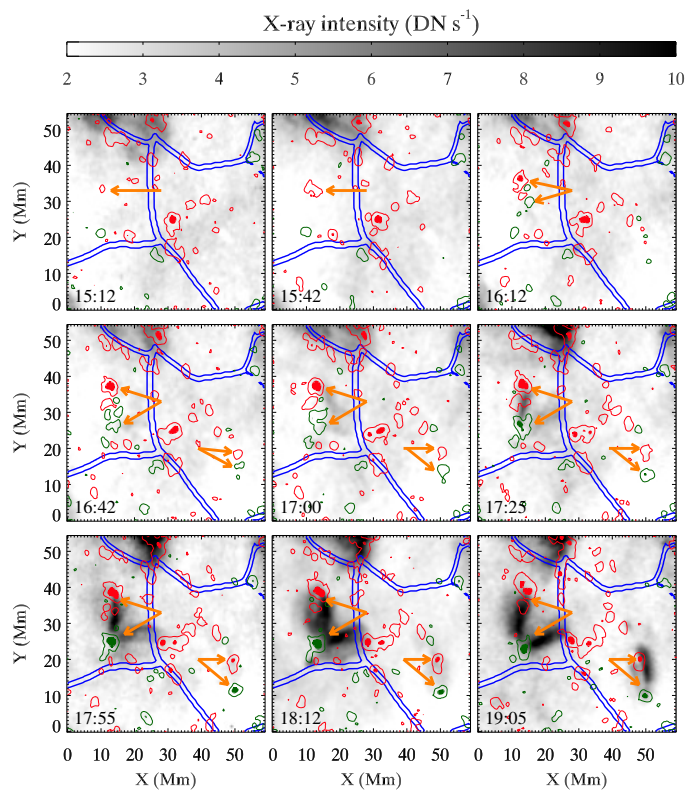}
	\caption{Flux emergence observed at two different places within the displayed field of view. Red/green contours are positive/negative flux (respectively). Thin unfilled contours are at $10\G$, filled contours are at $50\G$. The orange arrows point at emerging flux regions that are followed by the rise of X-ray loops. The blue contours are the contours of the supergranular boundaries. \label{flux_em}}.
\end{figure}

The flux emergence is associated with the rise of X-ray loops observed at the same location, and shown in \Fig{flux_em} (pointed to by the orange arrows). The snapshots are taken in a FOV of $60\times60\Mm^2$. Balltracking was used to derive the flow fields, and the associated supergranular network lanes are drawn as blue contours. These are obtained with the algorithm of automatic recognition of supergranular cells from \citet{Potts08}. 

At 15:12, the flux is barely visible in the internetwork, until it emerges as a very fragmented, mixed-polarity flux after 16:00 (left arrows). This occurred quite close to the supergranular boundaries (blue lanes in \Fig{flux_em}, only $10\Mm$ away from it, which is consistent with previous observations of flux emergence \citep{Wang88,Stangalini2014}. The clustered flux then drifts away, and is finally observed with one clearly visible X-ray loop, at 17:55. At 18:12 another X-ray loop seems to connect the negative-polarity footpoint from the left side of the network lane in the middle of the frame, to the positive one on the other side.
A second, weaker emergence is seen near the bottom right part of the snapshots, starting at 16:42, and also gave rise to X-ray loops, visible at 19:05. 
The time scale of these emergences is a few hours. The amount of flux in the emerged bipoles that are observed at the footpoints of the X-ray loops is measured with the magnetic balltracking. The results are presented in the next section.

\subsection{Results of the segmentation}

Magnetic balltracking is performed on the FOV of the snapshots in \Fig{flux_em} to track the features pointed to by the orange arrows. The results are used by the region-growing algorithm: as explained in \Section{sec:regiongrow}, each ball can act as an identifier of a whole magnetic feature, using the ball numbers as unique labels. This makes it easier to detect and isolate only the emerging flux. To do so, we simply associate the number of the balls that were tracking these features to the time series of flux that is integrated by region-growing. Here, we selected only the flux that had increased at the end of the time series by twice the standard deviation. This is sufficient to isolate the emerging flux seen in \Fig{flux_em}. At this stage, all the emerging flux is not associated with bright X-ray emission. Finally, we use the ball numbers to select precisely the patches that are emerging underneath the X-ray loops. This last selection is not an automated process as we need to know where the X-ray loops are located, and to identify the ball numbers by eye. 

\begin{figure}
	\centering
	\includegraphics[width=0.9\columnwidth]{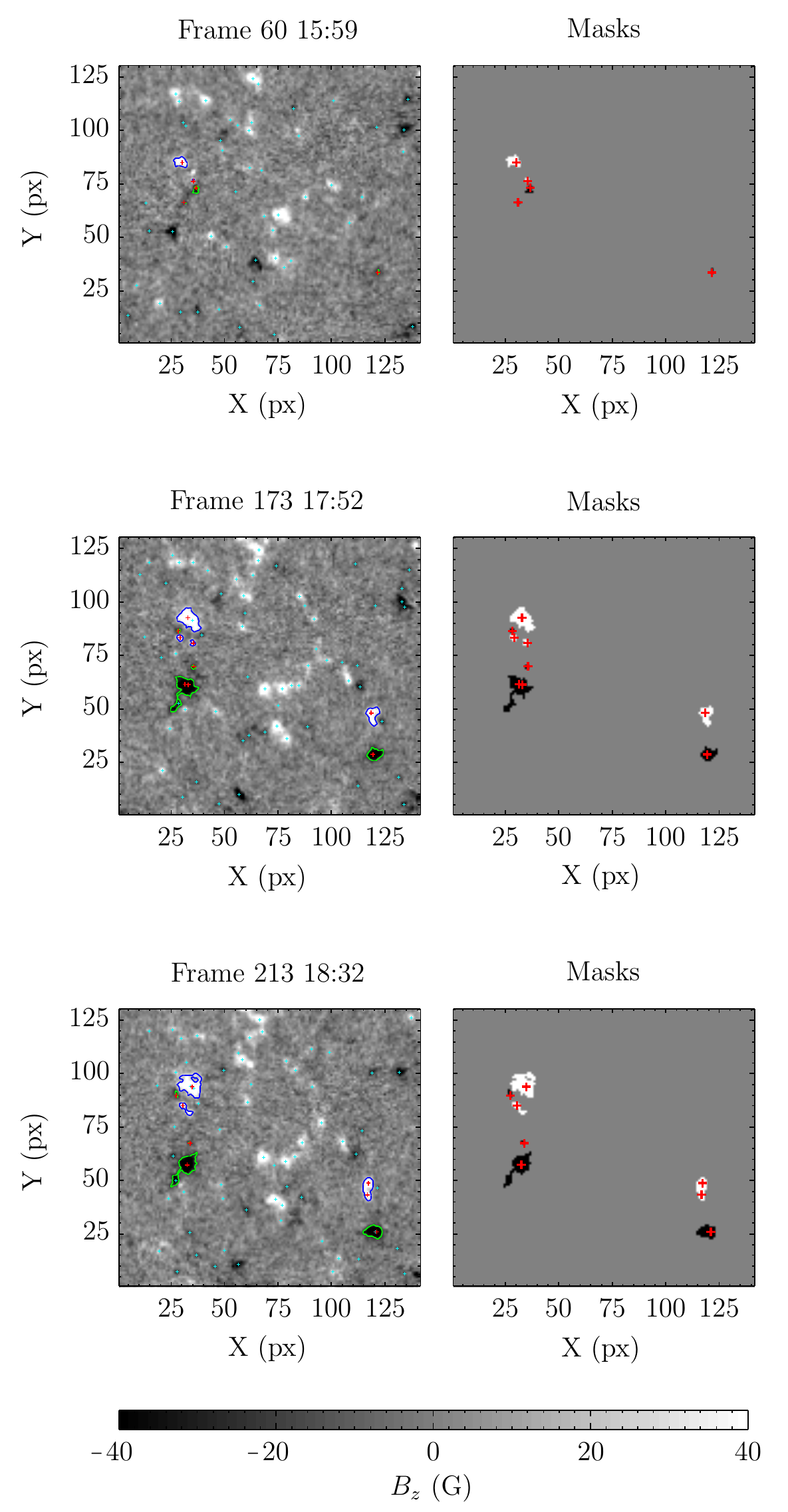} 
	\caption{Snapshot of the region extraction of the emerging flux. The right column are the magnetograms. The left column are the masks of the extracted regions. The contours are at $\pm 20\G$. A movie showing the temporal evolution is available in the online edition.}
	\label{snap2}
\end{figure} 

One can see the result of this selection more precisely in the three snapshots shown in \Fig{snap2}, and in the second movie (online supplemental). The left panel shows the magnetograms with the positions of the balls (red crosses) that are tracking the emerging flux (see also \Fig{snap2}). The small turquoise crosses show the positions of the balls used throughout the magnetic balltracking, excluding the ones added during Phase 4. The bigger red crosses correspond to the balls selected for the region extraction of the emerging flux. The blue and green contours outline the boundaries of the extracted masks that are used to spatially integrate the flux density in the magnetograms. These masks are shown as binary patches in the right panel (black for negative flux, white for positive flux). They are the main output of the region extraction algorithm (\S~\ref{sec:regiongrow}). The detection threshold was set to $20\G$, which is near the noise level of the magnetograms in MDI. The pixels with an intensity below this value are ignored. It is possible to see the limitations of such a low threshold in the left panel of the movie, by considering the green contours on the left, near Frames 173-174 and again around Frame 213. The extracted region suddenly includes small nearby features for 1 to 2 frames. This illustrates possible errors when using region growing with a detection threshold too close to the noise level. Nonetheless, the short time scale (1 to 2 frames) of such extraction errors, compared to the lifetime of the correctly extracted features (here, an order of magnitude greater) makes it possible to filter them out, for instance with median filtering over a few time steps, or using Fourier filtering in the time-frequency domain, without significantly impairing the rest of the data. However, these techniques are not part of the magnetic balltracking and we will not discuss them further. 

\begin{figure} 
	\centering
		\subfigure{\includegraphics[width=1\columnwidth]{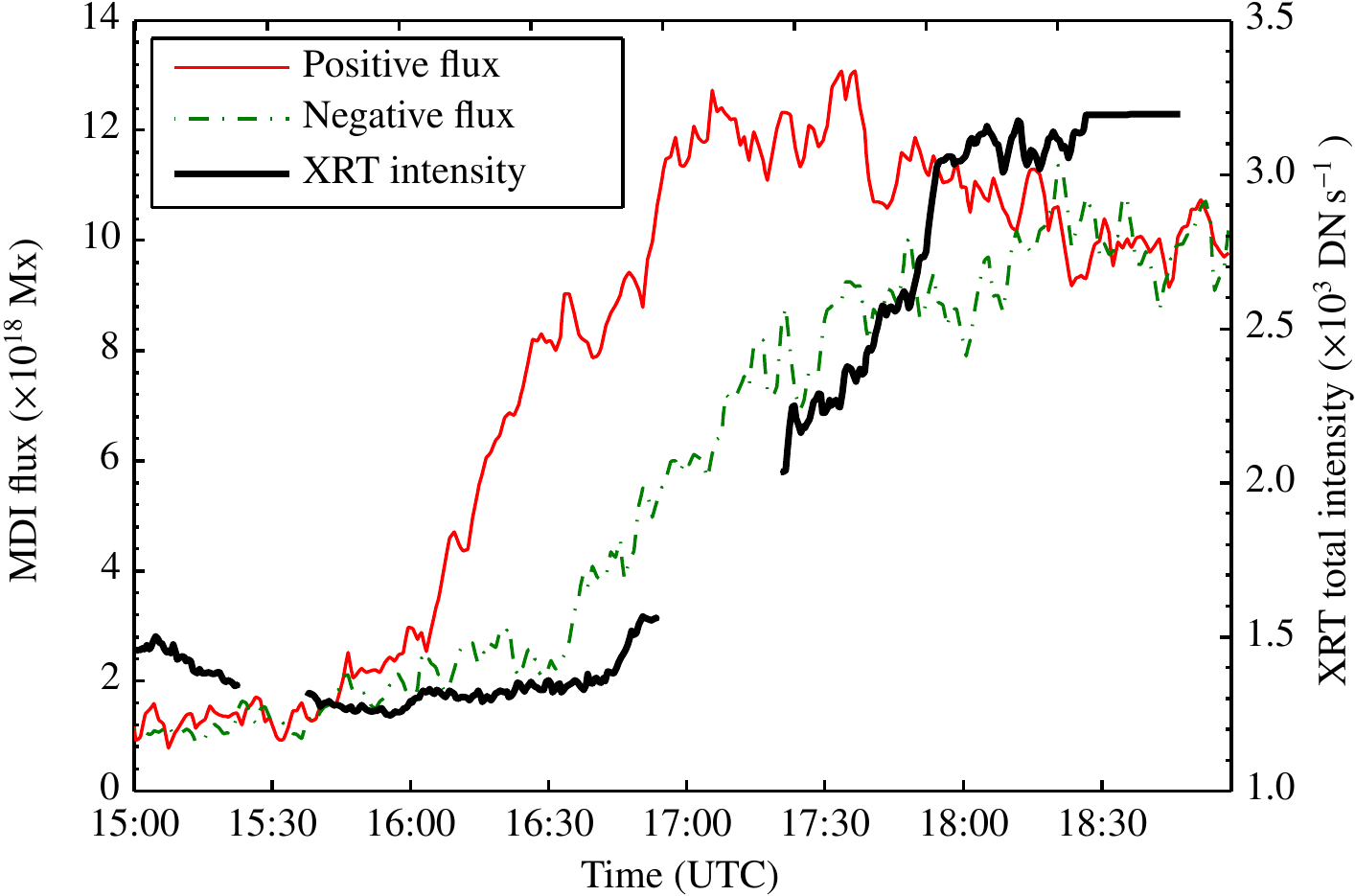}}
		\subfigure{\includegraphics[width=1\columnwidth]{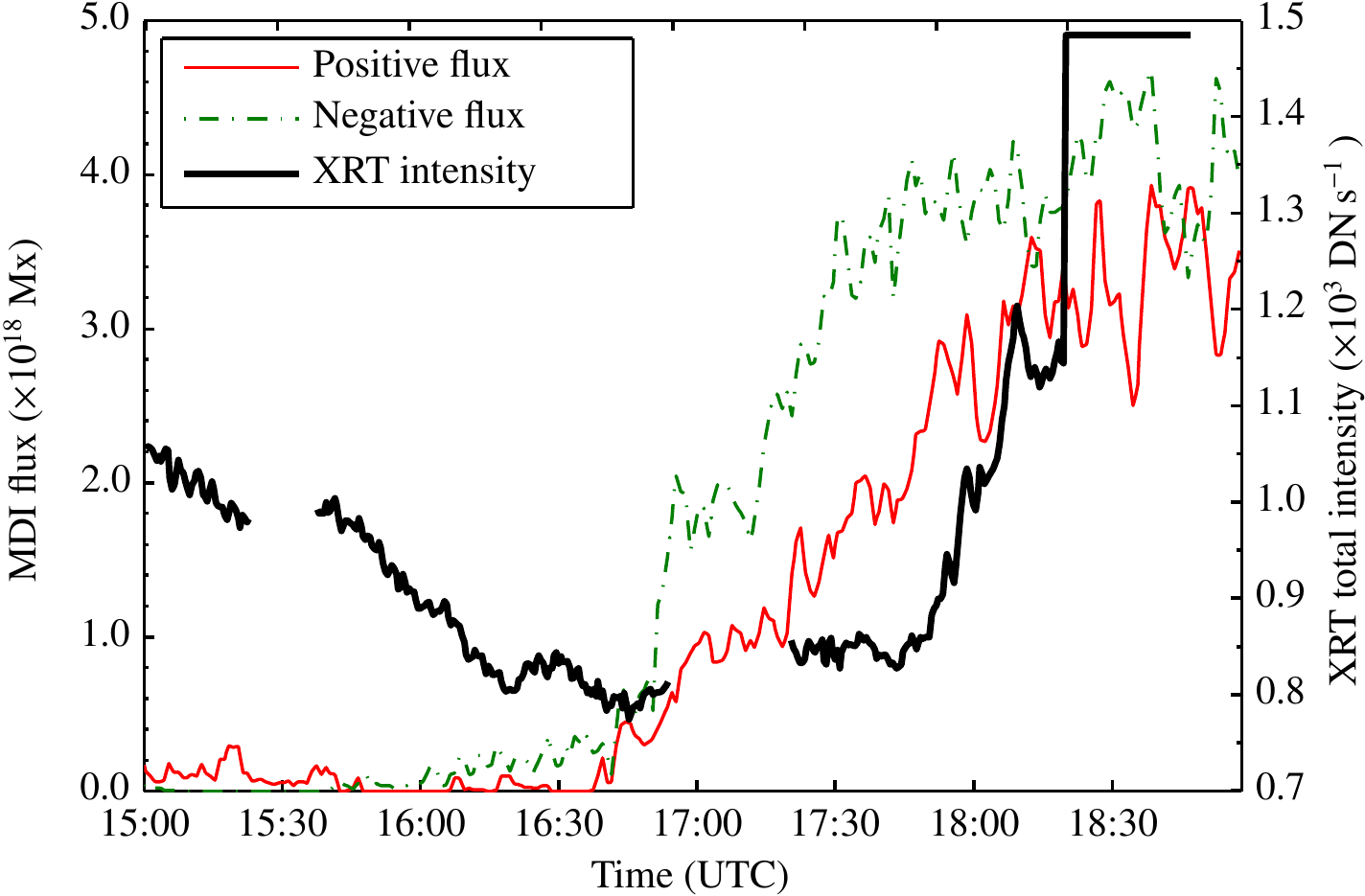}}
	\caption{Top: evolution of the X-ray intensity (black line) and the magnetic flux in the region of emerging flux on the left part of the snapshots in \Fig{flux_em}. The colours of the curves of the flux are consistent with the contours. Red is positive flux, green is negative flux (in absolute value). 
Bottom: Same as the top panel, for the second region of flux emergence in the bottom right of the snapshots in \Fig{flux_em}. There is an X-ray data gap between 18:40 and 19:02 that has been filled in by the first available value after the gap.}
	\label{emergence_flux}
\end{figure}

The integrated fluxes are plotted in \Fig{emergence_flux}. The flux in the first emergence (\Fig{flux_em}, left-side arrows, and \Fig{emergence_flux}, top) is balanced at the beginning (15:00), for $\simm40\minutes$ with a positive and a negative flux of a bit less than $2 \times 10^{18} \Mx$. It is unbalanced for $\simm2.5\hr$, until it is balanced again at $\sim$18:15 with an unsigned flux of $\simm10^{19} \Mx$. The X-ray emission increases $\simm1\hr$ after the emergence is first detected, by $\simm70\%$ of the background intensity from $\simm1200\DNS$ at 15:00 up to $\simm3200\DNS$\footnotemark[3] after 18:30. 
\footnotetext[3]{$\DNS$: Data Number per Second.}

In the second case of emergence (\Fig{flux_em}, right-side arrows, and \Fig{emergence_flux}, bottom), the magnetic flux is about 50\% weaker than in the first region, with a maximum positive and negative flux (respectively) between $3.5 \times 10^{18}\Mx$ and $4.5 \times 10^{18}\Mx$. The flux is unbalanced during about $2\hr$, between $\sim$16:40 and $\sim$18:40, although the flux balance is not obvious afterwards. Like in the previous case, this flux emergence is followed by the rise of an X-ray loop. The X-ray emission increases by about 50\% from a background level of $1000\DNS$ at 15:00, up to $\simm1500\DNS$ after 19:00 when the X-ray loop is visible (\Fig{flux_em}, bottom right panel). 
Note that there is an X-ray data gap between 18:36 and 19:02. The X-Ray gap is filled with the first value available after the gap (19:02), which is only an arbitrary cosmetic correction. 

\subsection*{Comment on the results}

The focus is given here on the current capability of the algorithm, regardless of the capability of the instrument. Therefore, for a better assessment of future improvements of the algorithm, this first example of a science application of magnetic balltracking is using data as close as possible to what is delivered by the instrument. Therefore, the magnetograms are not free of P-mode oscillations that may contaminate the Zeeman-related signals \citep{DeForest2007}. Here these oscillations directly affect the region-extraction algorithm by changing the apparent flux density and area of the extracted feature. The less the signal-to-noise ratio in the extracted area, the more visible the oscillations are. This is responsible for the relatively greater oscillations in the second case of the flux emergence in \Fig{emergence_flux} (bottom), which we observed to be 50\% weaker than in the first case (top). This patently shows that for actual statistical analyses with magnetic balltracking, the preprocessing guidelines in \citet{DeForest2007}, which are not, \textit{per se}, part of the magnetic balltracking, should be followed.

\section{Summary and prospects}
In this paper, we have presented our implementation of an efficient method called "magnetic balltracking" that tracks the magnetic features down to their finest scales. This new algorithm allows us to quantify the evolution of the magnetic flux. Applied on MDI data , it allowed us to detect, track, and quantify the evolution of emerging flux between $10^{18}\Mx$ to $10^{19}\Mx$ on a very fine scale of a few $\Mm$, and that are followed by the rise of soft X-ray loops within a few hours. 

Although we have presented here an application to estimate the flux over the tracked features, magnetic balltracking has a much broader range of possible applications. It could be applied to the tracking of the footpoints of magnetic field lines when performing extrapolations, the study of MHD waves in flux tubes, or the diffusion of internetwork elements whose emergence, transport and disappearance can now be tracked in detail. In addition, the technique may be applied to  the formation of Active Regions. Because magnetic balltracking can detect emerging flux and track the elements as they grow and move apart, it could complement other algorithms used in surveys of the solar cycle that use different algorithms to detect and track sunspots, e.g., \citet{Watson2009,Watson2011} and \citet{Goel2014}. 

\begin{acknowledgements}
This work was funded by the International Max Planck Research School in G\"ottingen (Germany), and by grant STFC/F002941/1 from the UK's Science and Technology Facilities Council, held at the School of Physics and Astronomy, University of Glasgow. 
\end{acknowledgements}

\bibliographystyle{aa}
\bibliography{./sun2}

\begin{thebibliography}{21}
\expandafter\ifx\csname natexlab\endcsname\relax\def\natexlab#1{#1}\fi

\bibitem[{{Attie} {et~al.}(2009){Attie}, {Innes}, \& {Potts}}]{Attie09}
{Attie}, R., {Innes}, D.~E., \& {Potts}, H.~E. 2009, A\&A, 493, L13

\bibitem[{{Chae} {et~al.}(2007){Chae}, {Moon}, {Park}, {Ichimoto}, {Sakurai},
  {Suematsu}, {Tsuneta}, {Katsukawa}, {Shimizu}, {Shine}, {Tarbell}, {Title},
  {Lites}, {Kubo}, {Nagata}, \& {Yokoyama}}]{Chae2007}
{Chae}, J., {Moon}, Y., {Park}, Y., {et~al.} 2007, PASJ, 59, 619

\bibitem[{{DeForest} {et~al.}(2007){DeForest}, {Hagenaar}, {Lamb}, {Parnell},
  \& {Welsch}}]{DeForest2007}
{DeForest}, C.~E., {Hagenaar}, H.~J., {Lamb}, D.~A., {Parnell}, C.~E., \&
  {Welsch}, B.~T. 2007, ApJ, 666, 576

\bibitem[{{Delaboudini{\`e}re} {et~al.}(1995){Delaboudini{\`e}re}, {Artzner},
  {Brunaud}, {Gabriel}, {Hochedez}, {Millier}, {Song}, {Au}, {Dere}, {Howard},
  {Kreplin}, {Michels}, {Moses}, {Defise}, {Jamar}, {Rochus}, {Chauvineau},
  {Marioge}, {Catura}, {Lemen}, {Shing}, {Stern}, {Gurman}, {Neupert},
  {Maucherat}, {Clette}, {Cugnon}, \& {van Dessel}}]{Delaboudiniere1995}
{Delaboudini{\`e}re}, J., {Artzner}, G.~E., {Brunaud}, J., {et~al.} 1995, Sol.
  Phys., 162, 291

\bibitem[{{D{\'e}moulin} \& {Berger}(2003)}]{Demoulin2003}
{D{\'e}moulin}, P. \& {Berger}, M.~A. 2003, Sol. Phys., 215, 203

\bibitem[{{Goel} \& {Mathew}(2014)}]{Goel2014}
{Goel}, S. \& {Mathew}, S.~K. 2014, Sol. Phys., 289, 1413

\bibitem[{{Golub} {et~al.}(2007){Golub}, {Deluca}, {Austin}, {Bookbinder},
  {Caldwell}, {Cheimets}, {Cirtain}, {Cosmo}, {Reid}, {Sette}, {Weber},
  {Sakao}, {Kano}, {Shibasaki}, {Hara}, {Tsuneta}, {Kumagai}, {Tamura},
  {Shimojo}, {McCracken}, {Carpenter}, {Haight}, {Siler}, {Wright}, {Tucker},
  {Rutledge}, {Barbera}, {Peres}, \& {Varisco}}]{Golub07}
{Golub}, L., {Deluca}, E., {Austin}, G., {et~al.} 2007, Sol. Phys., 243, 63

\bibitem[{Gonzalez \& Woods(2008)}]{Gonzalez2008digital}
Gonzalez, R. \& Woods, R. 2008, Digital Image Processing, 3rd edn.
  (Pearson/Prentice Hall)

\bibitem[{{Hagenaar} {et~al.}(1999){Hagenaar}, {Schrijver}, {Title}, \&
  {Shine}}]{Hagenaar1999}
{Hagenaar}, H.~J., {Schrijver}, C.~J., {Title}, A.~M., \& {Shine}, R.~A. 1999,
  ApJ, 511, 932

\bibitem[{{Lamb} {et~al.}(2008){Lamb}, {DeForest}, {Hagenaar}, {Parnell}, \&
  {Welsch}}]{Lamb2008}
{Lamb}, D.~A., {DeForest}, C.~E., {Hagenaar}, H.~J., {Parnell}, C.~E., \&
  {Welsch}, B.~T. 2008, ApJ, 674, 520

\bibitem[{{Parnell}(2002)}]{Parnell2002}
{Parnell}, C.~E. 2002, MNRAS, 335, 389

\bibitem[{{Potts} {et~al.}(2004){Potts}, {Barrett}, \& {Diver}}]{Potts04}
{Potts}, H.~E., {Barrett}, R.~K., \& {Diver}, D.~A. 2004, A\&A, 424, 253

\bibitem[{{Potts} \& {Diver}(2008)}]{Potts08}
{Potts}, H.~E. \& {Diver}, D.~A. 2008, Sol. Phys., 248, 263

\bibitem[{{Scherrer} {et~al.}(1995){Scherrer}, {Bogart}, {Bush}, {Hoeksema},
  {Kosovichev}, {Schou}, {Rosenberg}, {Springer}, {Tarbell}, {Title},
  {Wolfson}, {Zayer}, \& {MDI Engineering Team}}]{Setal95}
{Scherrer}, P.~H., {Bogart}, R.~S., {Bush}, R.~I., {et~al.} 1995, Sol. Phys.,
  162, 129

\bibitem[{{Stangalini}(2014)}]{Stangalini2014}
{Stangalini}, M. 2014, A\&A, 561, L6

\bibitem[{{Tsuneta} {et~al.}(2008){Tsuneta}, {Ichimoto}, {Katsukawa}, {Nagata},
  {Otsubo}, {Shimizu}, {Suematsu}, {Nakagiri}, {Noguchi}, {Tarbell}, {Title},
  {Shine}, {Rosenberg}, {Hoffmann}, {Jurcevich}, {Kushner}, {Levay}, {Lites},
  {Elmore}, {Matsushita}, {Kawaguchi}, {Saito}, {Mikami}, {Hill}, \&
  {Owens}}]{Tsuneta08}
{Tsuneta}, S., {Ichimoto}, K., {Katsukawa}, Y., {et~al.} 2008, Sol. Phys., 249,
  167

\bibitem[{{Wang}(1988)}]{Wang88}
{Wang}, H. 1988, Sol. Phys., 116, 1

\bibitem[{{Watson} {et~al.}(2009){Watson}, {Fletcher}, {Dalla}, \&
  {Marshall}}]{Watson2009}
{Watson}, F., {Fletcher}, L., {Dalla}, S., \& {Marshall}, S. 2009, Sol. Phys.,
  260, 5

\bibitem[{{Watson} {et~al.}(2011){Watson}, {Fletcher}, \&
  {Marshall}}]{Watson2011}
{Watson}, F.~T., {Fletcher}, L., \& {Marshall}, S. 2011, {\aap}, 533, A14

\bibitem[{{Welsch} {et~al.}(2004){Welsch}, {Fisher}, {Abbett}, \&
  {Regnier}}]{Welsch2004}
{Welsch}, B.~T., {Fisher}, G.~H., {Abbett}, W.~P., \& {Regnier}, S. 2004, ApJ,
  610, 1148

\bibitem[{{Welsch} \& {Longcope}(2003)}]{Welsch03}
{Welsch}, B.~T. \& {Longcope}, D.~W. 2003, ApJ, 588, 620

\end{thebibliography}

\end{document}